\documentclass[11pt,prd,superscriptaddress,nofootinbib]{revtex4}
\usepackage[english]{babel}
\usepackage{graphicx}
\usepackage{mathtext}
\usepackage{indentfirst}
\usepackage{epsfig,amsmath,amsfonts}
\newcommand{\beq}[1]{\begin{eqnarray}\label{#1}}
\newcommand\eeq {\end{eqnarray}}
\newcommand\bqa {\begin{eqnarray}}
\newcommand\eqa {\end{eqnarray}}

\newcommand{\eq}[1]{(\ref{#1})}
\newcommand{\bear}{\begin{array}}
\newcommand{\enar}{\end{array}}

\begin{document}

\title{\Large Classical radiation by free-falling charges in de Sitter spacetime}
%\title{\bf Classical Radiation of Massless Fields by Free Falling Charges
%in de Sitter Space}
\author{\bf E.\ T.\ Akhmedov}
%\email{akhmedov@itep.ru}
\affiliation{B.\ Cheremushkinskaya, 25, ITEP, 117218, Moscow, Russia}
\affiliation{Moscow Institute of Physics and Technology, Dolgoprudny, Russia}
\author{\bf Albert Roura}
%\email{}
\affiliation{Max-Planck-Institut f\"ur Gravitationsphysik (Albert-Einstein-Institut),
Am M\"uhlenberg 1, 14476 Golm, Germany}
\author{\bf A.\ Sadofyev}
%\email{sadofyev@itep.ru}
\affiliation{B.\ Cheremushkinskaya, 25, ITEP, 117218, Moscow, Russia}
\affiliation{Moscow Institute of Physics and Technology, Dolgoprudny, Russia}

%\date{}

\begin{abstract}
We study the classical radiation emitted by free-falling charges in de Sitter spacetime coupled to different kinds of fields. Specifically we consider the cases of the electromagnetic field, linearized gravity and scalar fields with arbitrary mass and curvature coupling. Given an arbitrary set of such charges, there is a generic result for sufficiently late times which corresponds to each charge being surrounded by a field zone with negligible influence from the other charges. Furthermore, we explicitly find a static solution in the static patch adapted to a charge (implying no energy loss by the charge) which can be regularly extended beyond the horizon to the full de Sitter spacetime, and show that any other solution decays at late times to this one. On the other hand, for non-conformal scalar fields the inertial observers naturally associated with spatially flat coordinates will see a non-vanishing flux far from the horizon, which will fall off more slowly than the inverse square of the distance for sufficiently light fields ($m^2 + \xi R < 5H^2/4$) and give rise to a total integrated flux that grows unboundedly with the radius. This can be qualitatively interpreted as a consequence of a classical parametric amplification of the field generated by the charge due to the time-dependent background spacetime.
Most of these results do not hold for massless minimally coupled scalar fields, whose special behavior is analyzed separately.
\end{abstract}

\maketitle

\section{Introduction.}

The study of interacting quantum field theories in de Sitter spacetime \cite{birrell94,Chernikov:1968zm,Mottola:1984ar, Myrhvold,Dolgov:1994cq,Tsamis:1992zt} has recently been the subject of an increasingly renewed interest \cite{Weinberg:2005vy,Polyakov:2007mm,Garriga:2007zk,PerezNadal:2008ju,PerezNadal:2009hr,Roura:1999fq,Higuchi:2008tn,Alvarez:2009kq,Akhmedov:2009ta,Burgess:2010dd,Marolf:2010zp,Antoniadis:2006wq,Akhmedova:2008dz,Volovik:2008ww}. There is controversy in the literature on the significance of large quantum infrared effects and whether they can lead to an instability of de Sitter-invariant states and even a secular screening of the cosmological constant. In this respect, it is good to have solid answers to basic questions concerning field theory in de Sitter space.

In this note we want to analyze in detail the possible classical radiation of scalar, electromagnetic and linearized gravitational fields by free-falling charges in de Sitter spacetime. The fact that a charge follows a geodesic does not necessarily imply the absence of radiation. This is clearly illustrated by the example of an electrically charged (or neutral) small mass orbiting around a central massive object in an asymptotically flat spacetime, which will emit electromagnetic (unless the orbiting mass is neutral) and gravitational radiation. In this case there is a natural set of asymptotic inertial observers with respect to which the orbiting mass describes a nontrivial motion, and the existence of radiation can be understood fairly naturally; the situation is, however, less clear for de Sitter. Note that the emission of radiation in the previous example will lead to a back-reaction force which will deviate the actual trajectory from the original geodesic motion. In contrast, the high degree of symmetry of de Sitter guarantees that the trajectory of a single free-falling charge will not be altered even if it radiates.

In flat space radiation is defined as a far-zone field which can be separated from the near-zone field associated with the sources \cite{LL}, and falls off slowly enough so that it can propagate the total energy flux (integrated over the whole solid angle) over an arbitrarily long distance (in the absence of absorption or damping). Thus, in formulating the concept of radiation as emission of energy by the charges and subsequent energy propagation independently of them, local and global energy conservation arguments play a crucial role. These can be understood as a consequence of Minkowski spacetime possessing globally timelike Killing vectors.

When considering radiation in curved spacetimes, one needs to confront two new conceptual difficulties. The first one is the absence of energy conservation. Even though the total stress tensor is covariantly conserved, one can no longer relate the change of the energy within a given region to the flux through the surface enclosing that region in a simple and direct way. This is still possible when there is a timelike Killing vector available (see Sec.~\ref{sec:killing_energy}), but this is not the case in general. It is, therefore, possible to introduce this kind of construction within a static patch of de Sitter, leading to a conserved Killing energy, but not globally.
(In asymptotically flat spacetimes it is also possible to define consistently a notion of total energy and asymptotic energy flux in terms of the ADM and Bondi masses, and related concepts \cite{wald84}.)

The second new difficulty is the absence in general of a preferred frame of observers with respect to which one could naturally define the energy density and local energy fluxes (given any frame characterized by a tetrad $\{e_a^\mu\}$, they are obtained by contracting the stress tensor with the tetrad appropriately).
%We already saw above the key role of energy conservation in the usual
%understanding of the notion of radiation in flat space.
The importance and possible sensitivity to the choice of frame of observers can
%also
be illustrated with a simple example in flat space which shows that the flux can be zero for one set of observers, but nonzero for a different set. Consider an eternally uniformly accelerated electric charge in Minkowski space (with proper linear acceleration $a$). An inertial observer (placed at a distance much greater than $1/a$ from the charge) would see a nonzero flux. On the other hand, Rindlerian observers ``comoving'' with the charge would find a vanishing flux  \cite{fulton60}. (The easiest way to see this is by realizing that in the Rindler frame one has a static charge which produces a static electric field \cite{fulton60}.)
%Note that the Rindlerian observer does not detect radiation
%despite that the invariant energy loss rate of the charge
%(the one proportional to the square of the 4--acceleration \cite{LL}) is not zero.

We will bear these issues in mind when studying the fields generated by free-falling charges in de Sitter. As a consequence, we will separately analyze the energy loss by the source and the fall-off of the integrated flux at large radii (larger than the horizon), two aspects which are independent due to lack of global energy conservation. Furthermore, we will examine each case in several different frames. On the one hand, we will consider the inertial observers naturally associated with spatially flat and global coordinates, which are both particularly suitable for studying fluxes at radii much larger than the horizon. (Working with global coordinates can unmask global restrictions, such as the requirement for a vanishing total electric charge, which are absent otherwise.) On the other hand, we will also study the problem in terms of static coordinates, whose associated static observers (except for the central one) are non-inertial. Within the static patch covered by these coordinates one can define a conserved Killing energy and they are particularly useful to analyze the energy loss by the charge (which must have its origin in the current-field interaction term).

Various aspects of classical fields and radiation of massless scalars in de Sitter space have been studied in a number of papers
\cite{Tsamis:1992xa,deVega,Poisson,Higuchi:2008fu} (certain quantum mechanical features of radiation emission in de Sitter have also been analyzed in Refs.~\cite{Akhmedov:2009be,Bros:2009bz}). Here we will address the question of radiation of free-falling charges in quite some generality. We consider a broad class of fields (scalar with arbitrary mass and curvature coupling, electromagnetic and linearized gravity) and obtain the generic late-time field for an arbitrary set of free-falling charges. Furthermore, we examine the situation in several different frames to get a more complete picture, and pay especial attention to those aspects connected to radiation, namely, energy loss by the charge and fluxes at large distances.

The paper is organized as follows.
We start by considering the electromagnetic and conformal scalar fields in Secs.~\ref{sec:e-m} and \ref{sec:conformal}. In both cases the calculation can be reduced to a calculation in flat space by using a suitable conformal transformation, and complete absence of energy flux is found.
Next, the massless minimally coupled scalar field is considered in Sec.~\ref{sec:minimal}. The charge is found to lose energy at a constant rate, but the total integrated flux falls off to zero at large radii.
The case of general mass and curvature coupling is studied in Sec.~\ref{sec:general_scalar} and solutions implying no energy loss by the charge are found for $M^2=m^2 +\xi R > 0$. Nevertheless, for sufficiently light fields one has non-vanishing total flux at large radii.
In Sec.~\ref{sec:gravity} we analyze the case of linearized gravity.
The generic late-time behavior for an arbitrary set of geodesics is described in Sec.~\ref{sec:late}. Finally, we discuss our results in Sec.~\ref{sec:discussion}. Further details about coordinate systems in de Sitter and the regular extension of the static solution to the full spacetime are provided in a couple of appendices.

We use the $(+,+,+)$ sign convention of Ref.~\cite{misner73} and work throughout the paper with natural units ($\hbar=c=1$) such that the Hubble constant equals one.

\section{Electromagnetic field}
\label{sec:e-m}

We start with the calculation of the electromagnetic field produced by a free-falling
charge (i.e.\ moving along a geodesic). Since any two geodesics in de Sitter can be related by an isometry transformation, we can consider without loss of generality the geodesic corresponding to $\theta_3=x^i=r=0$ in the different coordinate systems (global, spatially flat and static) described in Appendix~A. %Appendix~\ref{appA}.
When working with global coordinates, we will need to include explicitly a second particle with opposite charge, which will be considered to move along the antipodal geodesic.
Taking into account that Maxwell's theory is conformally invariant and de Sitter space is conformal to the Einstein static universe or to flat space (for the region covered by a spatially flat foliation), we expect no electromagnetic radiation to be emitted by a free-falling charge in de Sitter. We will see that explicitly in the next subsections.

The action for the electromagnetic field coupled to a current $J^\mu$ is
\begin{equation}
S[A_\mu,J^\mu] = \int d^4x \sqrt{-g} \left( \frac{1}{4} F_{\mu\nu}F^{\mu\nu}
+ A_\mu J^\mu \right) \label{eq:em_action},
\end{equation}
%where $F_{\mu\nu} = \nabla_{\mu} A_{\nu} - \nabla_{\nu} A_{\mu}
%= \partial_{\mu} A_{\nu} - \partial_{\nu} A_{\mu}$.
where $F_{\mu\nu} = 2\, \nabla_{[\mu} A_{\nu]} = 2\, \partial_{[\mu} A_{\nu]}$ and the indices are raised with the inverse metric as usual.
It leads to the following equation of motion:
\begin{equation}
\nabla_\mu F^{\mu\nu} = \frac{1}{\sqrt{-g}}  \partial_\mu (\sqrt{-g} F^{\mu\nu})
= J^\nu \label{eq:maxwell1},
\end{equation}
For a point-like charge with worldline $z^\mu (\tau)$ the current is given by
\begin{equation}
J^\mu(x) = q \int d\tau \, \frac{\dot{z^\mu} (\tau)}{\sqrt{-g}} \,
\delta^{(4)} \big(x^\alpha - z^\alpha(\tau)\big)
= q \frac{\dot{z^\mu} (\tau)}{\sqrt{g^{(3)}}} \,
\delta^{(3)} \big(x^\alpha - z^\alpha(\tau)\big) \label{eq:current1}.
\end{equation}
with $\dot{ } \equiv d/d\tau$ and
where the first equality is valid for any parameterization of the worldline, but the second one is only valid when $\tau$ is the proper time and (locally) synchronous coordinates adapted to the worldline are employed, so that $g^{(3)}_{\mu\nu}$ is the spatial metric of the subspace orthogonal to the worldline.
The stress tensor for the electromagnetic field, which can be obtained by functionally differentiating the action for the free field, is
\begin{equation}
T_{\mu\nu} = F_{\mu\alpha} F^{\ \alpha}_\nu
- \frac{1}{4} g_{\mu\nu} F_{\alpha\beta}F^{\alpha\beta} .
\end{equation}
As mentioned above, one can see that the action in eq.~\eq{eq:em_action} is invariant under a conformal transformation of the metric, $g_{\mu\nu} \to \Omega^2(x)\, g_{\mu\nu}$, while keeping $A_\mu$ (and hence $F_{\mu\nu}$) fixed. Note, however, that the physical values of the electric and magnetic fields, which corresponds to contracting $F_{\mu\nu}$ with an orthonormal basis $\{\vec{e}_a\}$ will transform as $F_{\mu\nu} e^\mu_a e^\nu_b \to \Omega^{-2} F_{\mu\nu} e^\mu_a e^\nu_b$ due to the rescaling required to keep the basis orthonormal. Similarly, although $T_{\mu\nu}$ would rescale with a factor $\Omega^{-2}$, the physical values of the energy density, pressure and energy flux will transform as $T_{\mu\nu} e^\mu_a e^\nu_b \to \Omega^{-4} T_{\mu\nu} e^\mu_a e^\nu_b$. Finally, from eq.~\eq{eq:current1} one can see that the current and its physical values transform respectively as follows: $J^\mu \to \Omega^{-4} J^\mu$ and $J^\mu e_\mu^a \to \Omega^{-3} J^\mu e_\mu^a$.

\subsection{Spatially flat coordinates}

As seen in eq.~\eq{eq:planar2}, when parametrizing half of de Sitter space using spatially flat coordinates, it is conformal to half of Minkowski spacetime with conformal factor $\Omega^2 = 1/\eta^2$. Therefore, the electromagnetic field for a free-falling charge with $x^i=\mathrm{const.}$ [and hence $dx^{\mu}/d\eta=(1,0,0,0)$] will be given in the Coulomb gauge by%
\footnote{Note that the Lorentz gauge condition $\nabla_\mu A^\mu = 0$, satisfied by our solution in Minkowski spacetime, is not preserved in general under conformal transformations.}
\begin{equation}
A_{0}=\frac{q}{4\pi |\vec{x}|}, \quad A_i=0
\quad \mathrm{and} \quad F_{0i} = q\frac{x^i}{4\pi |\vec{x}|^3} ,
\end{equation}
where $|\vec{x}| = (\delta_{ij}x^ix^j)^{1/2}$ and we only wrote the non-vanishing components of the field strength. The physical electric field, $F_{\hat{0}\hat{i}} = q (\eta^2/4\pi|\vec{x}|^2) (x^i/|\vec{x}|)$, points in the radial direction and its magnitude is inversely proportional to the square of the physical distance on spatial sections of constant time. Since there is no magnetic field in this frame, the energy flux measured by the inertial observers comoving with the charge is zero, i.e. $T_{0i} = 0$.

\subsection{Global coordinates}

The situation is very similar when considering the global coordinates described in Appendix~A.1. In this case, one can take advantage of the fact that the spacetime is manifestly conformal, with a conformal factor $a^2(\lambda) = \cos^{-2}(\lambda)$, to a region of the Einstein static universe with line element $ds^2= -d\lambda^2 + d\Omega_3^2$\, and $-\pi/2 < \lambda < \pi/2$.

Working in the Coulomb gauge, $\nabla_i A^i = 0 = \nabla_\mu A^\mu$, one can easily find the solution for a free-falling charge at $\theta_3 = 0$ in the Einstein static universe:
\begin{equation}
A_{0}= \frac{q}{4\pi} \cot(\theta_3), \quad A_i=0
\quad \mathrm{and} \quad F_{03} = \frac{q}{4\pi \sin^2(\theta_3)}
\label{eq:em_field2},
\end{equation}
where $F^{\mu\nu}$ can also be obtained directly from eq.~\eq{eq:maxwell1}.
Because of conformal symmetry, the solutions have exactly the same form in global de Sitter coordinates (although $A_\mu$ no longer satisfies the Lorentz gauge condition in de Sitter space). The physical electric field is given by $F_{\hat{0}\hat{3}} = q / 4\pi a^2(\lambda) \sin^2(\theta_3)$, which is inversely proportional to the physical area of the 2-sphere centered at the charge and contained on the 3-spheres corresponding to the constant-time spatial sections.
Note also that the solution in eq.~\eq{eq:em_field2} implies the existence of a charge with opposite sign at the antipodal geodesic ($\theta_3 = \pi$). This can be understood as a consequence of Gauss's law [the $0$ component of eq.~\eq{eq:maxwell1}], which implies that the total charge (i.e. the integral of the charge density over the compact spatial sections) should vanish.

As in the case of spatially flat coordinates, the energy flux seen by the comoving inertial observers in the global frame vanishes ($T_{0i}=0$) due to  the absence of a magnetic field.

\subsection{Static coordinates}

Working with the static coordinates defined in Appendix~A.3, one can also solve Maxwell's equations easily to obtain the following result in the Coulomb gauge ($\nabla_i A^i = 0 = \nabla_\mu A^\mu$):
\begin{equation}
A_t = \frac{q}{4\pi r}, \quad A_i=0
\quad \mathrm{and} \quad F_{tr} = \frac{q}{4\pi r^2}
\label{eq:em_field3},
\end{equation}
where the component $F_{tr}$ coincides in this case with physical value of the electric field corresponding to the component $F_{\hat{t}\hat{r}}$ in an orthonormal basis. Once again the absence of a magnetic field implies a vanishing energy flux seen by the static observers ($T_{tr} = 0$).

Note that the results for $F_{\mu\nu}$ that we have obtained in the three coordinate systems are equivalent. This can be immediately seen by using eqs.~\eq{eq:transf1}-\eq{eq:transf5} to show that the functional dependence of the only non-vanishing component of $F_{\hat{\mu}\hat{\nu}}$ is the same in all cases, and taking into account that the transformation which relates the orthonormal bases associated with the different coordinate systems leaves the relevant element of the exterior basis invariant (the transformation in tangent and cotangent space is a boost along the direction of the electric field, which leaves it invariant). The solution in static coordinates also agrees with the result for the electromagnetic field of the de Sitter-Reissner-Nordstrom solution in the limit of very small mass and charge.

\subsubsection{Killing-energy conservation}
\label{sec:killing_energy}

The existence of a timelike Killing vector $\xi^\mu$ in the static patch (whose tangent curves are parametrized by $t$) allows the definition of a conserved Killing energy constructed from the current $j_\mu^{(\xi)} = T_{\mu\nu} \xi^\nu$, which is covariantly conserved (i.e.\ $\nabla^\mu j_\mu^{(\xi)}$=0).
%[i.e. $\nabla^\mu (T_{\mu\nu} \xi^\nu) = 0$].
Applying Gauss's theorem to a cylindrical spacetime region with top and bottom boundaries corresponding to constant $t$ hypersurfaces plus a third boundary generated by time translation, and differentiating with respect to time, one gets
\begin{equation}
\frac{dE}{dt} \equiv
\frac{d}{dt} \int_{\Sigma_t} d^3 x \sqrt{g_{_\Sigma}} \, T_{\mu\nu}\, \xi^\nu \, n^\mu
= N \int_{\partial \Sigma_t} \! d^2 x \sqrt{g_{_{\partial \Sigma}}} \,
T_{\mu\nu}\, \xi^\nu \, k^\mu  \label{eq:ke_flux1},
\end{equation}
where $n^\mu$ is a the unit vector normal to the hypersurface $\Sigma_t$, $k^\mu$ is the outward unit vector normal to its boundary $\partial \Sigma_t$, and $N$ is the normalization factor of the Killing vector, defined by $\xi^\mu = N n^\mu$. This equation displays the conservation of the Killing energy by relating the rate of change of the energy contained within the surface $\partial \Sigma_t$ and the total flux crossing that surface. Note that in principle one needs to consider the total stress tensor because it was necessary to consider a covariantly conserved stress tensor ($\nabla^\mu T_{\mu\nu} = 0 $) when deriving eq.~\eqref{eq:ke_flux1}.

In particular one can apply this result to study the energy loss (emission) by the charge. In this paper we will focus on the contributions to the stress tensor from the field and current-field interaction term. (In the examples that we will be considering the trajectory of the free-falling charge will not be affected and the contribution of the free charge terms will not play a significant role.) As we will see, whenever one has a static solution for the field, the flux [the right-hand side of eq.~\eqref{eq:ke_flux1}] vanishes, and so does the contribution of the free field to the left-hand side. From this we can conclude that the charge did not experience any energy loss. And indeed the only changes to the charge energy (if its trajectory is not altered) can come from the current-field interaction term, which will remain constant for a static solution.
On the other hand, in Sec.~\ref{sec:minimal_static} we will have a non-stationary solution with a non-vanishing flux for a massless minimally coupled scalar field, and this kind of energy conservation argument %based on eq.~\eqref{eq:ke_flux1}
will provide an alternative way of obtaining the rate of change of the current-field interaction term.

%\subsection{Arbitrary geodesics and observers}

%The situation becomes a bit trickier when we consider more generic geodesics. By the
%de Sitter isometry we always can put one of the geodesics to coincide with $\theta_3=0$, but the
%other one will be more generic than $\theta_3=\pi$.
%The bad news is that generic geodesics
%do not transform into the geodesics after the conformal transformations.
%The good news is that after the transformation by an element of the de Sitter isometry
%group, which maps our special, $\theta_3=0$ or $\theta_3=\pi$, geodesic
%to a generic one, the above static field transforms into time dependent one, but
%without radiation zone. We explicitly check this for the scalar field below, where
%the formulas are simpler.

\section{Massless conformally coupled scalar field}
\label{sec:conformal}

As we will see, the situation for a massless conformally coupled scalar field is very similar to the electromagnetic case. The action for a real scalar field with arbitrary mass $m$, curvature-coupling parameter $\xi$ and coupled to an external source $J(x)$ is given by
\begin{equation}
S[\phi,J,g_{\mu\nu}] = -\int d^4x \sqrt{-g}
\left( \frac{1}{2} g^{\mu\nu}\, \partial_\mu \phi \, \partial_\nu\phi
+ \frac{1}{2} m^2 \phi^2 + \frac{1}{2} \xi R \phi^2  + \phi J \right)
\label{eq:scalar_action},
\end{equation}
from which one obtains the following equation of motion, known as the Klein-Gordon equation:
\begin{equation}
(\Box - m^2 -\xi R)\, \phi = \frac{1}{\sqrt{-g}}  \partial_\mu (\sqrt{-g} g^{\mu\nu}
\partial_\nu \phi) - (m^2 + \xi R)\, \phi= J
\label{eq:KG1},
\end{equation}
where $R$ is the Ricci scalar.
For a point-like charge with worldline $z^\mu (\tau)$ the current is given by
\begin{equation}
J(x) = q \int d\tau \, \frac{\sqrt{-\dot{z}^\mu \dot{z}_\mu}}{\sqrt{-g}} \,
\delta^{(4)} \big(x^\alpha - z^\alpha(\tau)\big)
= \frac{q}{\sqrt{g^{(3)}}} \,
\delta^{(3)} \big(x^\alpha - z^\alpha(\tau)\big) \label{eq:current2},
\end{equation}
where the first equality is valid for any parameterization of the worldline whereas the second one is only valid when %$\tau$ is the proper time and
(locally) synchronous coordinates adapted to the worldline are employed.
The stress tensor is obtained by functionally differentiating the action with respect to the metric \cite{birrell94}:
\begin{equation}
T_{\mu\nu} = \nabla_\mu \phi \nabla_\nu \phi
- \frac{1}{2} g_{\mu\nu} \left( \nabla^\alpha \phi \nabla_ \alpha \phi
+ m^2 \phi^2 \right)
+ \xi \left( g_{\mu\nu} \Box - \nabla_\mu \nabla_\nu + G_{\mu\nu}\right) \phi^2
%
%T_{\mu\nu} = (1-2\xi) \nabla_\mu \phi \nabla_\nu \phi
%- 2 \xi \phi \nabla_\mu \nabla_\nu \phi
%- \frac{1}{2} g_{\mu\nu} \Big( (1-4\xi) \nabla^\alpha \phi \nabla_ \alpha \phi
%- 4 \xi \phi \Box \phi + m^2 \phi^2 \Big)
%+ \xi G_{\mu\nu} \phi^2
\label{eq:stress2}.
\end{equation}

For vanishing mass and $\xi=1/6$ (commonly known as the conformal coupling case), the action $S[\phi,J,g_{\mu\nu}]$ is invariant (up to a total divergence) with respect to conformal transformations of the metric, $g_{\mu\nu} \to \bar{g}_{\mu\nu} = \Omega^2(x)\, g_{\mu\nu}$, together with the following rescalings of the field and the current: $\phi \to \bar{\phi} = \Omega^{-1} \phi$ and $J \to \bar{J} = \Omega^{-3} J$ [note that the current for a point-like charge, given by eq.~\eq{eq:current2}, does rescale appropriately]. It is then easy to see that the stress tensor transforms in a simple way under such conformal transformations:
\begin{equation}
\bar{T}_{\mu\nu} = \frac{2}{\sqrt{-\bar{g}}} \,
\frac{\delta S[\bar{\phi},\bar{J},\bar{g}_{\mu\nu}]}{\delta \bar{g}^{\mu\nu}}
= \frac{2 \Omega^{-2}}{\sqrt{-g}} \,
\frac{\delta S[\phi,J,g_{\mu\nu}]}{\delta g^{\mu\nu}}
= \Omega^{-2} T_{\mu\nu}
\label{eq:conformal2},
\end{equation}
where we used the invariance of the action with respect to conformal transformations in the second equality.
This simple transformation of the stress tensor under conformal rescalings will be exploited below to obtain the energy flux in de Sitter space from the result in Minkowski or the Einstein static universe.

\subsection{Spatially flat coordinates}

As mentioned above, for a massless conformally coupled field ($m=0,\ \xi=1/6$) the action is invariant under conformal transformations. Therefore, by taking $\Omega = -1/\eta$ one can immediately relate the field generated by a free-falling scalar charge in de Sitter when using spatially flat coordinates to the analogous situations in flat space, where the solution is given by $\phi = -q/4\pi |\vec{x}|$. The corresponding solution in de Sitter is then obtained by rescaling the field:
\begin{equation}
\phi = - \frac{q}{4 \pi} \frac{(-\eta)}{|\vec{x}|} \label{eq:hyperg2a}.
\end{equation}
Since the Minkowski solution is time independent, there is no energy flux associated with it ($T_{0i} = 0$). Eq.~\eq{eq:conformal2} then implies the absence of energy flux as seen by the comoving observers ($x^i = \mathrm{const.}$) in de Sitter spacetime with spatially flat coordinates.

\subsection{Global coordinates}

Similarly, the solution for a free-falling scalar charge at $\theta_3=0$ in global coordinates can be straightforwardly obtained by solving the Klein-Gordon equation in the conformally related Einstein static universe, where the relevant equation to solve (if one looks for a static solution) is
\begin{equation}
\frac{1}{\sin^2(\theta_3)}\frac{\partial}{\partial \theta_3} \left( \sin^2(\theta_3)
\frac{\partial \phi}{\partial \theta_3} \right) - \phi
= \frac{q}{4 \pi} \frac{\delta(\theta_3)}{\sin^2(\theta_3)}
\label{eq:einstein1},
\end{equation}
which follows from applying eq.~\eq{eq:KG1} with $m=0$ and $\xi=1/6)$ to the Einstein static universe and taking into account that $R=6$ when the radius of curvature is one. Introducing $z=\cos(\theta_3)$, eq.~\eq{eq:einstein1} becomes
\begin{equation}
(1-z^2)\frac{\partial^2 \phi}{\partial z^2} - 3z \frac{\partial \phi}{\partial z}  - \phi
= \frac{q}{4 \pi \sqrt{2}} \frac{\delta(z-1)}{\sqrt{1-z}} \label{eq:einstein2}.
\end{equation}
The homogeneous part of this equation corresponds to the hypergeometric differential equation with parameters $a=b=1$ and $c=3/2$. Thus, the solution to eqs.~\eq{eq:einstein1} and \eq{eq:einstein2} is
\begin{equation}
\phi(z) = - \frac{q}{4 \pi^2} \, F\Big(1,1;3/2;\frac{1+z}{2}\Big)
= - \frac{q}{4 \pi^2} \frac{\pi - \theta_3}{\sin(\pi-\theta_3)}
= - \frac{q}{4 \pi \sin(\theta_3)} \left( 1- \frac{\theta_3}{\pi}\right)
\label{eq:hyperg1},
\end{equation}
which is the only solution of the homogeneous equation which is regular everywhere except for $z=1$, where the source is located, and has the right behavior as $z\to1$ compatible with the right-hand side of eqs.~\eq{eq:einstein1} and \eq{eq:einstein2} [namely, $\phi \to -q/4\pi \theta_3$ as $\theta_3 \to 0$]. There is a second independent solution of the homogeneous equation,
\begin{equation}
\phi(z) = - \frac{q}{4 \pi^2} \left[ F\Big(1,1;3/2;\frac{1+z}{2}\Big)
+ F\Big(1,1;3/2;\frac{1-z}{2}\Big) \right]
= - \frac{q}{4 \pi \sin(\theta_3)}
\label{eq:hyperg2},
\end{equation}
which is not regular at $z=-1$ either and actually corresponds to having also an identical source at the antipodal worldline ($\theta_3=\pi$). This solution will play an important role when comparing with the solutions in the other subsections and in Sec.~\ref{sec:general_scalar}. Other solutions of the homogeneous equation can be obtained by taking a linear combination of eqs.~\eq{eq:hyperg1} and \eq{eq:hyperg2} and will correspond to a certain combination of sources at $\theta_3=0$ and $\theta_3=\pi$.

Note that in contrast with the electromagnetic case, where Gauss's law forced the total charge on the compact spatial section to be zero, even static solutions of the Klein-Gordon equation can have a non-vanishing total charge in the Einstein static universe due to the term proportional to the Ricci scalar and the field in eq.~\eq{eq:KG1}, as illustrated by the solution in eq.~\eq{eq:hyperg1}. In fact, by considering a continuous superposition of rotations on $S^3$ of that solution, one can generate a solution which corresponds to an arbitrary charge distribution on $S^3$.

The solutions in de Sitter space can be immediately obtained from those in the Einstein static universe dividing by $a(T) = \cosh(T)$. For instance, the solution for two identical antipodal charges becomes
\begin{equation}
\phi(T,\theta_3) = - \frac{q}{4 \pi \cosh(T) \sin(\theta_3)}
\label{eq:hyperg2b}.
\end{equation}
The static solutions in the Einstein static universe have obviously $T_{0i}=0$. Taking into account eq.~\eq{eq:conformal2}, we conclude that there is no energy flux as seen by the comoving observers (with $\theta^i=\mathrm{const.}$) in de Sitter global coordinates for the solution in eq.~\eq{eq:hyperg2b} or any solution obtained by rescaling the static solutions mentioned above for the Einstein static universe.

\subsection{Static coordinates}

Finally, one can work with the static coordinates of Appendix~A.3. If we consider a free-falling charge at $r=0$, the static solutions will satisfy the following equation
\begin{equation}
\frac{1}{r^2} \frac{\partial}{\partial r}
\left( r^2 (1-r^2) \frac{\partial \phi}{\partial r} \right)
- 2 \phi = \frac{q}{4 \pi} \frac{\delta(r)}{r^2} \label{eq:static4}.
\end{equation}
Unfortunately in this case it is not straightforward to find explicit solutions in terms of known special functions, but one can still study their existence and their main properties. This will be done in Sec.~\ref{sec:general_scalar} for arbitrary values of $\xi$ and $m$. As we will see, eq.~\eq{eq:static4} has a one-parameter family of solutions, but only one of them is regular on the horizon (at $r=1$). From the construction in Sec.~\ref{sec:general_scalar} and Appendix~B, it is clear that such a solution should correspond to eq.~\eq{eq:hyperg2b}, which is regular everywhere except for the charge worldlines at $\theta_3=0$ and $\theta_3=\pi$. Using eq.~\eq{eq:transf3} it can be written in static coordinates as
\begin{equation}
\phi(t,r) = - \frac{q}{4 \pi r}
\label{eq:hyperg2c},
\end{equation}
which can be easily checked to satisfy eq.~\eq{eq:static4}. Using eq.~\eq{eq:transf1} one can also see that it is equivalent to the solution in spatially flat coordinates given by eq.~\eq{eq:hyperg2a}. Being time-independent, this solution has zero flux as seen by the static observers with $r=\mathrm{const.}$ (i.e. $T_{tr}=0$).

\section{Massless minimally coupled scalar field}
\label{sec:minimal}

The main features of the massless minimally coupled case are the absence of a stationary solution in the static patch which can be regularly extended to the full spacetime, and the existence, instead, of solutions which exhibit a constant non-vanishing energy flux as seen by static observers. Some of the results presented here have previously been obtained in Ref.~\cite{Poisson}, where this problem was considered.

\subsection{Spatially flat coordinates}
\label{sec:minimal_flat}

In this case conformal symmetry is no longer available and we need to solve explicitly eq.~\eqref{eq:KG1} with $m=\xi=0$, which becomes
(after multiplying by $\sqrt{-g}$)
\begin{equation}
\partial_{\mu}\frac{\eta^{\mu\nu}}{\eta^2}\partial_{\nu}\phi
= \frac{q}{\eta}\,\delta^{(3)}(\vec{x}) \label{eq:KG5},
\end{equation}
where we particularized the current in eq.~\eqref{eq:current2} to a free-falling point-like charge located at $x^i=0$.
The retarded Green function for a massless minimally coupled scalar field in de Sitter spacetime is a well known result. An explicit derivation starting from eq.~\eqref{eq:KG5} can be found for instance in Ref.~\cite{Poisson} and the result is
\begin{equation}
G_\mathrm{R}(\eta,\vec{x};\,\eta',\vec{x}') = \left[\frac{\eta\,\eta'}{4\pi|\vec{x}-\vec{x}'|} \, \delta\left(\eta- \eta' - \left|\vec{x} - \vec{x}'\right|\right) + \frac{1}{4\pi}\theta\left(\eta - \eta' - \left|\vec{x} - \vec{x}'\right|\right)\right]\, \theta\left(\eta - \eta'\right),
\end{equation}
where one should remember that $\eta$ is negative and future infinity corresponds to $\eta=0$. Hence, we will have the following solution of eq.~(\ref{eq:KG5}):
\begin{equation}
\phi = q \int d^4x' \,
G_\mathrm{R}(\eta,\vec{x};\eta',\vec{x}')\,\frac{\delta^{(3)}(\vec{x}')}{\eta'}
= \frac{q}{4\pi} \left[\frac{\eta}{|\vec{x}|}
+\log\left(\frac{|\eta - |\vec{x}||}{|\eta_{0}|}\right)\right] + \mathrm{const.}
\label{eq:radsol}
\end{equation}
where $\eta_0$ is an arbitrary dimensionful parameter and the remaining constant term is actually divergent and reflects the logarithmic divergence exhibited by the integral with respect to $\eta'$. Nevertheless, one can simply discard this constant term and it will still be a solution of eq.~\eqref{eq:KG5}. Note that in contrast with Ref.~\cite{Poisson}, our solution is valid for all times, although it becomes arbitrarily large at early times. This point and the possibility of ``turning on'' the charge at a finite initial time $\eta_\mathrm{i}$, as done in Ref.~\cite{Poisson}, will be further discussed in Sec.~\ref{sec:minimal_static}.

Calculating the time and spatial derivatives of the solution found in eq.~\eqref{eq:radsol},
\bqa
\partial_{0}\phi &=& \frac{q}{4\pi} \left(\frac{1}{|\vec{x}|}+\frac{1}{\eta - |\vec{x}|}\right) , \label{radsol1a} \\
\partial_{i}\phi &=& \frac{q}{4\pi} \left(-\frac{\eta\, x_i}{|\vec{x}|^3}+\frac{x_i}{|\vec{x}|\,(|\vec{x}| - \eta)}\right) ,
\label{radsol1b}
\eqa
we can find the energy flux, whose components in an orthonormal basis are given by $T_{\hat{0}\hat{\imath}} = \eta^2 (\partial_{0}\phi) (\partial_{i}\phi)$.
This flux is invariant under the isometry associated with the time translation in eq.~\eqref{eq:isometry1}. The first contribution to $\phi$, $\partial_0 \phi$ and $\partial_{i} \phi$ dominates for points well within the horizon ($|\vec{x}| \ll |\eta|$) and coincides with the solution for the massless conformally coupled case. Its contribution to $T_{\hat{0}\hat{\imath}}$ is not zero, but falls off fast enough so that its total integrated flux vanishes in the limit of infinite radius. (In the conformal case $T_{\hat{0}\hat{\imath}}$ vanishes because there is an exact cancellation from the $\xi \,\nabla_\mu\nabla_\nu\phi^2$ term.)
For points outside the horizon ($|\vec{x}| > |\eta|$) the second terms on the right-hand side of eqs.~\eqref{radsol1a}-\eqref{radsol1b} must also be taken into account. Such a term dominates then the contribution to $\partial_i \phi$ for large $|\vec{x}|$ and is inversely proportional to $|\vec{x}|$. On the other hand, the two terms on the right right-hand side of eq.~\eqref{radsol1a} become comparable and partially cancel out, so that $\partial_{0}\phi$ falls off like $1/|\vec{x}|^2$ for large $|\vec{x}|$. This means that the total integrated flux tends to zero for large radii. [These points can be seen more directly %, and could even have been easily anticipated,
by considering first a large distance expansion on the right-hand side of eq.~\eqref{radsol1a}.]

\subsection{Static coordinates}
\label{sec:minimal_static}

%For the beginning let us stress that inflationary planar
%coordinates with $\tau = - \log(-\eta)$ correspond to the
%inertial observers, because in this reference frame coordinate time
%($\tau$ not $\eta$) coincides
%with the proper one. At the same time static coordinates are seen by non--inertial
%observers. However static coordinates have time--like Killing vector which allows
%to define energy, but it is not globally defined.

In this subsection we will consider the static coordinates of Appendix~A.3, but it will be convenient to introduce the tortoise radial coordinate $r_*$, which arises from requiring $dr_*/dr = (1-r^2)^{-1}$ and is related to the standard radial coordinate by $r=\tanh r_*$. In terms of this new coordinate, the line element becomes
\bqa
ds^2 = \frac{-dt^2 + dr_*^2}{\cosh^2(r_*)} + \tanh^2(r_*)\, d\Omega_2^2.
\eqa
We will consider the same free-falling source as in Sec.~\ref{sec:minimal_flat}, whose trajectory is given in terms of these coordinates by $x^\mu (t) = (t,0,0,0)$. The Klein-Gordon equation (including the source) for a spherically symmetric field will be
\bqa
\cosh(r_*)\,\left[- \tanh^2(r_*)\,\partial_t^2 \phi +
\partial_{r_*} \! \left( \tanh^2(r_*)\,\partial_{r_*} \phi \right) \right]
= \frac{q}{4\pi} \delta(r_*)
\label{eq:KG_tortoise1}.
\eqa
If we are interested in finding static solutions, the equation further reduces to
\bqa
\cosh^2(r_*)\, \partial^2_{r_*} \phi + 2\,\coth(r_*)\, \partial_{r_*} \phi =
\frac{q}{4\pi} \frac{\cosh^3(r_*)}{\sinh^2(r_*)}\, \delta(r_*)
= \frac{q}{4\pi\, r_*^2}\, \delta(r_*)
\label{eq:KG_tortoise2}.
\eqa
It is easy to verify that $\phi = - (q/4\pi)\left[\coth(r_*) - r_*\right]$ is a solution of this equation. Thus, it may naively seem that even for a massless minimally coupled field a free-falling charge in de Sitter can produce a stationary field, which implies $T_{0i}=0$, in its associated static patch. There are, however, several important additional aspects that should be taken into account. Let us start by pointing out that eq.~\eqref{eq:KG_tortoise1} has the following obvious time dependent solution
\bqa
\phi = - \frac{q}{4\pi} \left[\coth(r_*) - r_* + a\,t + b\right]
\label{statsol} ,
\eqa
for any real constants $a$ and $b$. Next, one can use eq.~\eqref{eq:transf1} as well as eq.~\eqref{eq:transf2} divided by eq.~\eqref{eq:transf1} to obtain the following relations between static and spatially flat coordinates:
\bqa
- \eta &=& \cosh(r_*)\, e^{-t},\nonumber \\
|\vec{x}| &=& \sinh(r_*)\, e^{-t} .
\eqa
Taking these into account, one can easily see that eqs.~\eqref{statsol} and \eqref{eq:radsol} are equivalent (after discarding the divergent constant piece in the latter) if one takes $a=1$ and $b = \log |\eta_0|$. But what is the explanation for the additional freedom in choosing $a$ and $b$ that one appears to have when working with static coordinates? This can be understood as follows. If we introduce the advanced and retarded times $v = t + r_*$ and $u = t - r_*$, we have $- r_* + a\,t = v\, (a-1)/2 - u\, (a+1)/2$, and we can clearly see that it will diverge on the future event horizon (which corresponds to $v \to \infty$ while keeping $u$ finite) unless $a=1$. Therefore, requiring that the solution is regular on the future event horizon and can be extended beyond it, completely fixes the existing freedom (up to the trivial additivity constant) and selects that solution already found in Sec.~\ref{sec:minimal_flat}, which is not stationary in the static patch. There is no stationary solution that can be regularly extended beyond the event horizon, but as we will see in the next section, this is a peculiarity of fields with $M^2=m^2 +\xi R = 0$.

Let us take the solution in eq.~\eqref{statsol} with $a=1$ and study the Killing energy and Killing energy fluxes along the lines presented in Sec.~\ref{sec:killing_energy}. Even though the solution is not stationary, it gives a time independent result for the $T_{tt}$ component of the field stress tensor an its contribution to the first integral in eq.~\eqref{statsol}. Taking this into account and using the same kind of arguments that led to the derivation of eq.~\eqref{eq:ke_flux1}, we can infer the following two conclusions. First, the total energy flux going through surfaces of different radii should be the same and, second, it should equal the energy loss by the charge (coming from the current-field interaction term). The first point can indeed be checked if one takes $N=1/\cosh(r_*),\ \sqrt{g_{\partial\Sigma}}=\tanh^2(r_*),\ T_{tr_*}= - (q/4\pi)^2 \coth^2(r_*),\ \xi^\mu=\delta^\mu_t,\ k^\nu=\cosh(r_*)\, \delta^\nu_{r_*}$ and substitutes them into the right-hand side of eq.~\eqref{eq:ke_flux1} to get a constant integrated flux (independent of the radius of the $\partial\Sigma_t$ surface) equal to $q^2/4\pi$. Furthermore, one can also check that the decrease in the total Killing energy, $dE/dt = - q^2/4\pi$, that follows from eq.~\eqref{eq:ke_flux1} for such an energy flux does agree with the decrease of Killing energy associated with the current-field interaction term, which is given by $q\, \partial \phi(t,0)/\partial t = - q^2/4\pi$.

Several additional remarks are in order. First, if we had included the standard kinetic term for the trajectory of the point-like charge with a mass $m$, we would have found (even allowing for a general trajectory of the charge) that the current-field interaction term can be interpreted as a contribution to the ``effective mass'' of the charge, given by $m_\mathrm{eff}(\tau) = m  +  q\, \phi\big(t(\tau),0\big)$. Note that the field diverges at the origin and one needs to introduce a suitable regularization and renormalization procedure, which involves taking an infinite bare mass $m$ that cancels the corresponding divergence (all this is necessary because we consider the strictly point-like particle limit). The important point is that within this context this procedure can be carried out in a time-independent manner, so that changes with time of the field will directly reflect changes of the effective mass; see Ref.~\cite{Poisson} for further details, but keep in mind that we use the opposite sign for the current-field interaction term here.
The second remark is that the solution in eq.~\eqref{statsol} with $a=1$ diverges on the past event horizon (this is easily seen in terms of $u,v$ coordinates, where the past horizon corresponds to $u \to -\infty$ while keeping $v$ finite), which is equivalent to the fact that the solution found in Sec.~\ref{sec:minimal_flat} diverged for $\eta \to -\infty$. This issue was not present in Ref.~\cite{Poisson}, where the charge was turned on at some finite time $t_0$, and it can be understood as follows: if the charge loses energy at a constant rate and one demands the mass to be bounded, it cannot be radiating for an infinite period of time. Hence, it was assumed in Ref.~\cite{Poisson} that the charge was turned on at some initial time $t_0$, when the renormalized effective mass had a finite value $m_0$. At later times, the effective mass would be given by $m_\mathrm{eff}(t) = m_0 - q^2/4\pi (t-t_0)$. We can see that the effective mass will vanish at a certain time and become negative after that. In Ref.~\cite{Poisson} it was essentially concluded that the charge would simply disappear after that. In our opinion, however, the question is not so clear. Think of a free-falling charge which was also electrically charged. It would not radiate electromagnetically, but it would lose energy due to the emission of the minimally coupled field. At some point its mass would vanish, but it cannot simply disappear because of electric charge conservation.
A nontrivial process is likely to happen at this point, but the actual details may be rather subtle and complex. We will return briefly to this question in the discussion section.

\subsection {Global coordinates}

Here we reanalyze the situation using global coordinates. We will only consider a single charge (this aspect will be further discussed below). The source on the right-hand side of the Klein-Gordon equation will then be given by $(q/4\pi)\,  \delta(\theta_3)/\sin^2(\theta_3) \cosh^3(T)$. There is a nice expression for the retarded propagator in terms of the invariant interval for a pair of points \cite{Allen:1985wd}:
\begin{equation}
G_\mathrm{R}(Z) = \frac{1}{4\pi} \big[\delta (1-Z)
+ \theta(1-Z)\big] \times \theta(T-T') ,
\end{equation}
where $Z(x,x')=X^A(x) X^B(x') \, \eta_{AB}$ is the de Sitter-invariant interval associated with the pair of points $x$ and $x'$. In terms of the coordinates introduced in Appendix~A.1 and for the particular configuration with $\theta'_3=0$ considered here, we have $Z(x,x') = \sinh(T) \sinh(T') - \cosh(T) \cosh(T')\cos(\theta_3)$.
Integrating with the source, we get the following intermediate expression for the solution
\bqa
\phi(T,\theta_3) = \frac{q}{4\pi}\,\int_{-\infty}^T \! dT' && \!\!\!\!
\Bigg[ \, \delta\Big(1 +  \sinh(T)\,\sinh(T') - \cosh(T)\,\cosh(T')\,\cos(\theta_3)\Big) \nonumber \\
&&+ \, \theta\Big(1 + \sinh(T)\,\sinh(T') - \cosh(T)\,\cosh(T')\,\cos(\theta_3)\Big)\Bigg] \label{eq:global_int}.
\eqa
The argument of the delta and theta functions is zero when
\bqa
e^{T'} = \frac{-1 \pm \cosh(T) \, \sin(\theta_3)}{\sinh(T) - \cosh(T)\,\cos(\theta_3)}.\label{lightcone}
\eqa
where the upper sign corresponds to the retarded time ($T'<T$) and the lower sign to the advanced one ($T'>T$). Moreover, there will be no solution when $\lambda<\theta_3-\pi/2$ (where $\lambda$ is the conformal time introduced in Appendix~A.1), which corresponds to spacetime points that lie outside the causal future of the entire charge worldline. In the end one gets the following final result:
\bqa
\phi(T,\theta_3) = \Bigg\{\!\!\!\!\!\!&&-\frac{q}{4\pi \cosh(T)\,\sin(\theta_3)}
\nonumber\\
&&+ \frac{q}{4\pi} \log \left[ \frac{- 1 + \cosh(T) \, \sin(\theta_3)}{\sinh(T) - \cosh(T)\,\cos(\theta_3)}\right] + \mathrm{const.} \Bigg\} \, \theta(\lambda-\theta_3+\pi/2)
\label{eq:global_sol},
\eqa
where the same remarks given below eq.~\eqref{eq:radsol} concerning the existence of a logarithmic divergence in the integration and the possibility of adding an arbitrary constant also apply here. In fact, it is considerably less cumbersome to evaluate the integral in eq.~\eqref{eq:global_int} by changing to spatially flat coordinates (this can be done because the support of the integrand is entirely contained within the region covered by the spatially flat coordinates), and the calculation becomes then very similar to that of Sec.~\ref{sec:minimal_flat}. Alternatively one can directly check that, when expressed in terms of spatially flat coordinates, eq.~\eqref{eq:global_sol} reduces to eq.~\eqref{eq:radsol}.
In order to show this equivalence one needs to make use of a couple of relations between spatially flat and global coordiantes: $-1/\eta = \sinh(T)  + \cosh(T) \cos(\theta_3)$, which follows from comparing the expressions for $X^0-X^4$ in Eqs.~\eqref{eq:global0} and \eqref{eq:planar1}, and $-|\vec{x}|/\eta = \cosh(T) \sin(\theta_3)$, which follows from comparing the expressions for $(X^iX^j\delta_{ij})^{1/2}$ in Eqs.~\eqref{eq:global0} and \eqref{eq:planar1}. In addition, one needs ot use the fact that the right-hand side of Eq.~\eqref{lightcone} with the upper choice of sign is identically equivalent to $(\sinh(T)  + \cosh(T) \cos(\theta_3))/(1+\cosh(T) \sin(\theta_3))$.

We close this subsection with several remarks. First, the solution in eq.~\eqref{eq:global_sol} diverges at $\lambda = \theta_3-\pi/2$, which corresponds to the divergence at $\eta \to -\infty$ or the past event horizon already found in Secs.~\ref{sec:minimal_flat}-\ref{sec:minimal_static} respectively. As discussed in Sec.~\ref{sec:minimal_static}, this can be cured by switching on the charge at a finite initial time (it can be done smoothly to guarantee that the Klein-Gordon equation is satisfied in the whole spacetime). In contrast with the electromagnetic or gravitational case, this is mathematically consistent because there is no symmetry that requires the charge to be conserved.
Second, note that it was possible to find a solution for the full de Sitter spacetime even though we considered a single charge. While the Gauss constraint forbids this possibility in the electromagnetic case and something similar happens also in the gravitational case (as briefly explained in Sec.~\ref{sec:gravity_global}), there is no such restriction for a scalar field. If we looked for time-independent solutions, $\nabla_i \nabla^i \phi = 0$ would imply a vanishing total charge in the massless minimally coupled case, but the existence of extra terms involving time derivatives in the Klein-Gordon equation means that there is no restriction on the total charge when one allows for a general time dependence of the solution.
On the other hand, demanding certain properties from the solution may sometimes imply the need to consider additional charges. For instance, we will see in the next section that having an identical antipodal charge makes it possible to have a stationary solution in the corresponding two static patches which can be regularly extended to the full de Sitter spacetime for $m^2 + \xi R > 0$. However, as we will see, that is not possible for $m^2 + \xi R = 0$, and there was consequently no particular motivation for considering more than one charge here.

\section{Scalar field with arbitrary mass and curvature coupling}
\label{sec:general_scalar}

We have seen that, similarly to the electromagnetic case, one can have a free-falling charge coupled to a conformal scalar field in de Sitter spacetime which does not radiate. In contrast, for a massless minimally coupled scalar field it is not possible to have a stationary solution in the static patch which is regular on the horizon (and can be extended beyond it). A charge coupled to such a field keeps losing energy indefinitely at a constant rate. We would like to investigate how generic this situation is: is the absence of radiation a peculiarity of the conformal case closely tied to conformal symmetry, or is on the contrary the situation for the massless minimally coupled case exceptional? (The case of vanishing mass and minimal coupling is known to exhibit several other peculiarities due to large infrared effects.) We will show in this section that the generic situation is the absence of energy loss by the charge. On the other hand, we will see in Sec.~\ref{sec:general_scalar2} that for sufficiently light fields the integrated energy flux seen by inertial observers at distances much larger than the horizon can grow unboundedly with the radius. The existence of these two seemingly contradictory results is a consequence of the lack of global energy conservation in de Sitter spacetime; see Sec.~\ref{sec:discussion} for a more detailed discussion.

More specifically, given a scalar field with arbitrary mass $m$ and curvature coupling $\xi$, whose action and equation of motion are listed in eqs.~\eq{eq:scalar_action}-\eq{eq:current2}, its dynamics for a fixed background spacetime with constant scalar curvature depends on the ``effective mass'' $M^2 = m^2 + \xi R$. We will see that given two identical antipodal charges at $\theta_3=0$ and $\theta_3=\pi$, there is always a globally regular solution for $M^2 > 0$ which is stationary in the corresponding static patches. This solution exhibits a vanishing energy flux as seen by the static observers. (Note that the antipodal source is entirely absent from the spacetime region covered by the spatially flat coordinates or a single static patch.)
%natural comoving observers associated with the global, spatially flat and static
%coordinate systems (the antipodal source is entirely absent from the spacetime
%region covered in the last two cases).

It is worth pointing out that having $M^2 > 0$ does not necessarily imply that the fields cannot exhibit massless behavior: it is not trivial to define unambiguously the notion of mass when the Compton wavelength is comparable or larger than the radius of curvature. In fact, the case that shares some key properties with massless fields in flat space (namely, a retarded propagator with support entirely contained on the light-cone for conformally flat spacetimes such as de Sitter) is the conformal case, which corresponds to $M^2 = 2$ (where we have used that $R=12$ in units with $H=1$).

%Although this paper is mostly dedicated to the issue of the
%radiation of only massless fields, in this section we would like to study the
%classical solutions for the massive scalars due to the free falling
%charges in de Sitter space. Note that one can not draw any conclusion about
%the radiation of massive fields via the calculation of the flux through a distant surface.
%Due to the mass of the radiation field such a flux will be always vanishing. To study
%the classical radiation of the massive scalars one should use a different approach \cite{Akhmedov:2009be}.
%In that paper it was shown, via calculation
%of the classical amplitude/intensity, that there is a radiation of the massive scalars
%from free falling particles as seen by inertial planar observers.

\subsection{Static coordinates}
\label{sec:general_scalar1}

The Klein-Gordon equation in static coordinates with a point-like source at $r=0$ is
\begin{equation}
-\frac{1}{1-r^2} \frac{\partial^2 \phi}{\partial t^2}
+ \frac{1}{r^2} \frac{\partial}{\partial r}
\left( r^2 (1-r^2) \frac{\partial \phi}{\partial r} \right)
+ \frac{\Delta_2 \phi}{r^2}
- M^2 \phi = \frac{q}{4 \pi} \frac{\delta(r)}{r^2} \label{eq:static5},
\end{equation}
where $\Delta_2$ is the Laplacian on the 2-sphere, covered with the coordinates $\{\theta_1,\theta_2\}$. If we look for static and spherically symmetric solutions centered at $r=0$, the equation reduces to
\begin{equation}
(1-r^2) \frac{\partial^2 \phi}{\partial r^2}
+ \frac{2}{r}(1-2r^2) \frac{\partial \phi}{\partial r}
- M^2 \phi = \frac{q}{4 \pi} \frac{\delta(r)}{r^2} \label{eq:static6}.
\end{equation}
The homogeneous part is a second-order ordinary differential equation with regular singular points at $r=0$, $r=1$ and $r=\infty$. Thus, Fuchs's theorem guarantees that one can use the Frobenius method to construct two independent solutions as a couple of convergent series.
The two solutions around $r=0$ can be chosen to be
\begin{eqnarray}
\phi_1(r) &=& \sum_{n=0}^\infty a_n r^n = 1 +\frac{M^2}{6} r^2 + \cdots \,,
\label{eq:frobenius1a}\\
\phi_2(r) &=& \frac{1}{r} \sum_{n=0}^\infty b_n r^n + c \, \ln(r) \, \phi_1(r)
= \frac{1}{r} + \left(\frac{M^2-2}{2}\right) r + \cdots  \label{eq:frobenius1b}\,.
\end{eqnarray}
The first one is regular at $r=0$ since $(\partial \phi_1/\partial r)(0)=0$ (otherwise it would have a spike at the origin of spherical coordinates and would not be differentiable there) and is, in fact, an even function of $r$. It is, therefore, a solution of the homogeneous equation with no source at the origin. On the other hand, the second solution (in this case $c=0$) corresponds to a Dirac delta source at the origin. With the right normalization, namely $\phi(r) = - (q/4\pi)\, \phi_2(r)$, it is a solution of eq.~\eq{eq:static6} including the point-like source.

We are interested in solutions which are regular on the horizon. So let us study the series expansion for the solutions around $r=1$ and then discuss how they match the solutions found above. In this case the two independent solutions can be chosen as
\begin{eqnarray}
\tilde{\phi}_1(r) &=& \sum_{n=0}^\infty a_n (r-1)^n
= 1 -\frac{M^2}{2} (r-1) + \cdots \,, \label{eq:frobenius2a}\\
\tilde{\phi}_2(r) &=& \sum_{n=1}^\infty b_n (r-1)^n + \ln(r-1) \, \tilde{\phi}_1(r-1)
\nonumber \\
&=& \ln(r-1) - \frac{M^2}{2} (r-1)\ln(r-1) + \left(\frac{M^2}{2} -4\right) (r-1) + \cdots \,.
\label{eq:frobenius2b}
\end{eqnarray}
The second solution is singular at $r=1$. Hence, we should consider a solution proportional to $\tilde{\phi}_1(r)$, which is finite and regular at $r=1$. In fact, it can be extended smoothly to the full de Sitter space, as discussed in detail in Appendix~B. In general, this solution will match a linear combination of the solutions found above: $C \tilde{\phi}_1(r) = A \phi_1(r) + B \phi_2(r)$. Taking $B= -(q/4\pi)$, so that it is a solution of eq.~\eq{eq:static6} including the delta source, completely fixes $A$ and $C$. Note that one should exclude the possibility that $B=0$ because this would imply that a solution of the equation with the source at $r=0$ would always pick a contribution proportional to $\tilde{\phi}_2(r)$ and would have singular behavior at $r=1$. In Appendix~B we prove that this possibility can always be excluded for $M^2>0$.

{F}rom the previous paragraph we can conclude that for $M^2>0$ one can always find a static solution $\phi_\mathrm{s}(r)$ associated with a free-falling charge in de Sitter which is regular on the horizon and can be extended to the full spacetime, as discussed in more detail in Appendix~B and the next two subsections.%
\footnote{In fact, one can consider $r>1$ in eq.~\eq{eq:static6} to study the extension of the solution to the remaining two non-static patches and introduce an analogous expansion in terms of $1/r$ around $r=\infty$, which is also a regular singular point. In that way, one obtains the same kind of late-time decay that will be described in the next two subsections, as well as the large-distance behavior for scales much larger than the horizon; see Appendix~B.2 for further details.}
Since it is time-independent in static coordinates, this solution exhibits zero energy flux as seen by the static observers ($T_{ti}=0$).
%as well as zero Killing energy flux, which is simply related to the former by a
%radially dependent redshift factor.

We close this subsection by briefly mentioning what things would have looked like if we had used the tortoise coordinate $r_*$ introduced in Sec.~\ref{sec:minimal_static}. The Klein-Gordon equation would be the same as eqs.~\eq{eq:KG_tortoise1} and \eq{eq:KG_tortoise2} with an extra $M^2\phi$ term on the left-hand side. Close to the source at $r=0$, $r$ and $r_*$ are almost the same and the solutions in terms of $r_*$ have the same form as in eqs.~\eq{eq:frobenius1a}-\eq{eq:frobenius1b}. On the other hand, close to the horizon, which corresponds to $r_* \to \infty$, the Klein-Gordon equation becomes approximately $\partial^2 \phi / \partial r_*^2 = 0$, whose solutions are a linear combination of $\bar{\phi}_1(r_*)=1$ and $\bar{\phi}_2(r_*)=r_*/2$. These correspond to the solutions in eqs.~\eq{eq:frobenius2a}-\eq{eq:frobenius2b}, and from this point on the argument would proceed exactly as above. In addition, one could study how perturbations with respect to this solution which are initially regular on the horizon decay at late times. This amounts to imposing the boundary condition of vanishing flux from $r_* \to \infty$ and is analogous to the quasinormal-mode analysis of perturbations outside a Schwarzschild black hole. We will, however, study the decay of perturbations using spatially flat and global coordinates.

\subsection{Spatially flat coordinates}
\label{sec:general_scalar2}

In spatially flat coordinates the Klein-Gordon equation for a field with an effective mass $M^2$ and a point-like source at $x^i=0$ is
\begin{equation}
- \frac{\partial^2 \phi}{\partial \tau^2}
- 3 H \frac{\partial \phi}{\partial \tau}
+ e^{-2\tau} \delta^{ij} \frac{\partial^2 \phi}{\partial x^i \partial x^j}
- M^2 \phi = q\, \delta^{(3)}(\vec{x})
\label{eq:flat1}.
\end{equation}
Given a solution $\phi_\mathrm{s}(r)$ in static coordinates like the one found in Sec.~\ref{sec:general_scalar1} (and its extension to $r>1$), one can explicitly check that $\phi_\mathrm{s}\big(e^\tau |\vec{x}| \big)$ is a solution of eq.~\eq{eq:flat1}, as it should be. Furthermore, its large-distance behavior can be obtained from eqs.~\eq{eq:frobenius3a}-\eq{eq:frobenius3b} and is given by
\begin{equation}
\phi_\mathrm{s}(\tau,x^i) \sim \Big( e^\tau |\vec{x}| \Big)^{-\lambda_-}
\label{eq:flat2},
\end{equation}
with $\lambda_- = 3/2 - \sqrt{9/4-M^2}$.
%From this expression one can straightforwardly derive the result for the
%energy flux at large $|\vec{x}|$.

Interestingly enough, one can establish the generic late-time behavior of an arbitrary solution and show that it tends at sufficiently late times to $\phi_\mathrm{s}\big(e^\tau |\vec{x}| \big)$. This follows from the fact that for $M^2>0$ the homogeneous solutions of eq.~\eq{eq:flat1} decay exponentially at late times. Indeed, if we Fourier transform the comoving spatial coordinates, the homogenous part of the Klein-Gordon equation becomes
\begin{equation}
\frac{\partial^2 \phi_k}{\partial \tau^2}
+ 3 H \frac{\partial^2 \phi_k}{\partial \tau}
+ \left( \vec{k}^2 e^{-2\tau} + M^2 \right) \phi_k = 0
\label{eq:flat3}.
\end{equation}
After a particular mode leaves the horizon (i.e. when $|\vec{k}| e^\tau > 1$) one can neglect the contribution from the term proportional to $\vec{k}^2$ in order to study the evolution of that mode. The solution at sufficiently late times is, thus, a linear combination of two exponentially decaying terms, $\exp(-\lambda_\pm \tau)$ with $\lambda_\pm = 3/2 \pm \sqrt{9/4-M^2}$, which always have a positive real part for $M^2>0$. In contrast, for $M^2=0$ we have $\lambda_-=0$ and one of the two solutions becomes constant outside the horizon. This can give rise to large infrared effects as modes keep leaving the horizon and is closely related to the peculiar behavior that we found for the massless minimally coupled case, including the absence of a regular static solution in the static patch.

Taking into account the explanations above, one can see that the approximate result for $\phi_\mathrm{s}(\tau,x^i)$ in eq.~\eq{eq:flat2} can be employed to calculate the energy flux at sufficiently late times and sufficiently large physical distances from the source. Its time and spatial derivatives are given by
\begin{eqnarray}
\frac{\partial \phi_\mathrm{s}}{\partial \tau}
&\sim& (- \lambda_-) \Big( e^\tau |\vec{x}| \Big)^{-\lambda_-}
\label{eq:flat4a}, \\
e^{-\tau} \frac{\partial \phi_\mathrm{s}}{\partial x^i}
&\sim& (- \lambda_-) \Big( e^\tau |\vec{x}| \Big)^{-1-\lambda_-}
\left(\frac{x^i}{|\vec{x}|}\right)
\label{eq:flat4b} .
\end{eqnarray}
The energy flux can then be obtained from eq.~\eq{eq:stress2}. Let us consider first the minimally coupled case ($\xi=0$). Only the first term on the right-hand side of eq.~\eq{eq:stress2} will contribute to the energy flux. At sufficiently late times and sufficiently far from the source, the result for the radial flux when working with an orthonormal basis is the following:
\begin{equation}
T_{\hat{0}\hat{r}} \sim (\lambda_-)^2 \Big( e^\tau |\vec{x}| \Big)^{-1-2\lambda_-}
\label{eq:flat5}.
\end{equation}
For positive but sufficiently small masses ($0<M^2 \ll 1$), $\lambda_-$ will be positive but very small. This means that the amplitude of the stress tensor will be quadratically suppressed, but, more importantly, it will fall off almost as slowly as the inverse of the physical distance. Therefore, the total flux integrated over a sphere of radius $R$ will grow almost proportionally to $R$ for sufficiently large radii.
This is a remarkable result whose interpretation and discussion is provided in Sec.~\ref{sec:discussion}.
In fact, in order to have a non-vanishing integrated flux at large radii it is sufficient to have $M^2 \leq 5/4$, which corresponds to $\lambda_- \leq 1/2$ ($M^2 = 5/4$ would give rise to a finite integrated flux in the limit of an infinite radius, whereas for smaller masses it would grow unboundedly with the radius).
For non-minimal coupling ($\xi \neq 0$)  there is an additional contribution from the term $\xi\, \nabla_\mu \nabla_\nu \phi^2$ in eq.~\eq{eq:stress2}, but the conclusions remain unchanged.

%@@ Comment on the terms that we are neglecting. @@

\subsection{Global coordinates}

In global coordinates the Klein-Gordon equation has the following form:
\begin{equation}
- \frac{\partial^2 \phi}{\partial T^2}
+ \frac{\Delta_3\, \phi}{\cosh^2 T} - M^2 \phi
= \frac{q}{4 \pi} \frac{\delta(\theta_3)}{\theta_3^2}
+ \frac{q}{4 \pi} \frac{\delta(\theta_3 - \pi)}{(\theta_3-\pi)^2}
\label{eq:global3},
\end{equation}
where $\Delta_3$ is the Laplacian of the 3-sphere, parametrized by $\{\theta_1,\theta_2,\theta_3\}$, and we included a free-falling source at $\theta_3=0$ as well as an antipodal one at $\theta_3=\pi$, which is necessary in order to have a static solution in the static patch which is globally regular. This is discussed in Appendix~B, where we also study its late-time and large-distance behavior, which is generically found to be
\begin{equation}
\phi_\mathrm{s}(T,\theta_3) \sim \Big( \cosh(T) \sin(\theta_3) \Big)^{-\lambda_-}
\label{eq:global4},
\end{equation}
and is valid for $\cosh(T) \sin(\theta_3) \gg 1$.
Note that at sufficiently late times, as the sources get farther and farther way, one will have non-negligible values of the field in two disconnected regions around each one of the sources with very similar form to that found above for spatially flat coordinates. As will be discussed in Sec.~\ref{sec:late}, this can actually be generalized to an arbitrary set of geodesics.

Similarly to the case of spatially flat coordinates one can also establish that the generic late-time behavior of an arbitrary solution of eq.~\eq{eq:global3} tends at sufficiently late times to $\phi_\mathrm{s}(T,\theta_3)$. The homogeneous solutions of eq.~\eq{eq:global3} can be decomposed in terms of
%a linear combination of
spherical harmonics $\mathcal{Y}_{LM}(\Omega)$ for the 3-sphere with time-dependent coefficients which are proportional to $(\cosh T)^{-3/2}$ times a linear combination of the Legendre functions $P_{L+1/2}^{\lambda_+ - 3/2}(\tanh T)$ and $Q_{L+1/2}^{\lambda_+ - 3/2}(\tanh T)$,
and decay at late times in the same way already found above for spatially flat coordinates.

%%$P_\nu^\mu(z) \sim \mathrm{const.} (1-z)^{-\mu/2}$ for $z$ close to $1$;
%%therefore, $P_{L+1/2}^{\lambda_+ - 3/2}(\tanh T) \sim \exp\big(-(\lambda_+
%%- 3/2)T\big)$, and the solution goes like $\exp(-\lambda_+ T)$.\\
%%$Q_\nu^\mu(z) \sim \mathrm{const.} (1-z)^{-\mu/2}
%%+ \mathrm{const.} (1-z)^{\mu/2}$ for $z$ close to $1$;
%%therefore, the solution goes like $\exp(-\lambda_+ T)$ plus
%%$\exp(-\lambda_- T)$.

The worldline of the antipodal source lies entirely outside the spacetime region covered by the spatially flat coordinates considered in Sec.~\ref{sec:general_scalar2}. We close this subsection briefly discussing its implications for the results found there. When solving eq.~\eqref{eq:flat1}, the effects of the antipodal source would enter through the initial value data specified on one of the flat spatial sections.
%(which could in turn be obtained by evolving the field in the full de Sitter
%spacetime, and would acquire the dependence on the antipodal source in the
%process).
At sufficiently late times any solution of eq.~\eqref{eq:flat1} will decay to $\phi_\mathrm{s}\big(e^\tau |\vec{x}| \big)$, which has no information left on the antipodal source. There is actually a simple interpretation of this point. Suppose that we start with $\theta_3 < \pi$ for the second source and then take the limit $\theta_3 \to \pi$.
Doing so while considering a flat spatial section with constant $\tau$, means that the location of the second source will be taken to an infinite physical distance on that section. Since the field generated by the second source will fall off at large distances in the same way as given by eq.~\eqref{eq:flat2},
%[see Sec.~\ref{sec:late} for a more general discussion involving an arbitrary
%number of geodesics],
it will not contribute in the limit of infinite separation (at least for $M^2>0$). That explains why there is no contribution to $\phi_\mathrm{s}\big(e^\tau |\vec{x}| \big)$ from the antipodal source.

\section{Linearized gravity}
\label{sec:gravity}

Gravitational radiation can be studied in the context of linearized gravity, regarded in this case as small metric perturbations around a de Sitter background geometry: $\tilde{g}_{\mu\nu}=g_{\mu\nu}+h_{\mu\nu}$. One can then evaluate the usual geometrical objects for the perturbed metric, expand in terms of the metric perturbation and retain the terms strictly linear in $h_{\mu\nu}$. It is convenient to express them as
\bqa
\Gamma^{\mu\, (1)}_{\nu \alpha} &=& \frac{1}{2} \big(\nabla_\alpha h^{\mu}_{\nu}
+ \nabla_\nu h^{\mu}_{\alpha} - \nabla^\mu h_{\nu \alpha} \big),\nonumber \\
R^{(1)}_{\mu\nu} &=& \frac{1}{2} \big( \nabla_\alpha \nabla_\nu h^{\alpha}_\mu + \nabla_ \alpha \nabla_ \mu h^{\alpha}_{\nu} - \nabla_ \nu \nabla_ \mu h
- \nabla^\alpha \nabla_\alpha h_{\mu\nu} \big) \label{eq:lin_ricci1},
\eqa
where $h\equiv g^{\alpha\beta} h_{\alpha\beta}$, $\nabla_\mu$ is the covariant derivative associated with the background metric $g_{\mu\nu}$, and we will be using the background metric to raise and lower indices unless explicitly stated otherwise.
An additional conceptual difficulty in this case is the existence of a gauge symmetry corresponding to transformations under diffeomorphisms. One can employ the background field method, as we did above, which preserves general covariance in terms of the background geometry. Nevertheless, the gauge freedom under transformations of the form $h_{\mu\nu} \to h_{\mu\nu} + 2 \nabla_{(\mu} \xi_{\nu)}$ needs to be fixed (it can be interpreted as the ambiguity in splitting the background and the perturbation). Even if one does so by imposing a gauge-fixing condition compatible with the background general covariance (background field gauge), in general the results will still depend on the particular gauge fixing. Only truly gauge-invariant physical observables will be completely independent.

In order to study gravitational radiation, we will follow Ref.~\cite{weinberg72} and compute $G_{\nu}^{\mu\, (2)}$, which results from evaluating the Einstein tensor for the perturbed metric $\tilde{g}_{\mu\nu}$ and keeping the terms strictly quadratic in $h_{\mu\nu}$, in a particular gauge. This object suffers from the gauge dependence mentioned above: while being generally covariant with respect to the background geometry, it will depend on the particular gauge which was chosen for $h_{\mu\nu}$.
(In asymptotically flat spacetimes it can still be employed to calculate gauge-invariant quantities such as the ADM mass, the total angular momentum, or the asymptotic energy flux, but that is not the case here.)
Notwithstanding, we will follow this common approach and leave deeper scrutiny for future work. In this respect it will likely be useful to consider the linearized Weyl tensor $C^{(1)}_{\mu\nu\rho\sigma}$, which is gauge invariant because its unperturbed counterpart vanishes and which will completely and naturally characterize the perturbed geometry at linear order.%
\footnote{Together with the linearized Weyl tensor, the linearized Riemann tensor is entirely determined by the linearized Ricci tensor, which can be cast in a gauge invariant form by raising one of the indices (the background quantity is then proportional to $\delta_\mu^\nu$, whose Lie derivative with respect to any vector vanishes). When one imposes the Einstein equation, this quantity vanishes  outside the sources: $R^{\nu\, (1)}_{\mu} = 0$.}
It might then be worth exploring the use of the Bel-Robinson super-energy tensor \cite{superenergy}, which involves terms quadratic in the Weyl, as a possible way of characterizing linearized gravitational radiation in a gauge invariant fashion.

\subsection{Spatially flat coordinates}

The linearized Einstein equation for metric perturbations around a de Sitter background can be written as
\begin{equation}
G^{\nu\, (1)}_{\mu} = 8 \pi G \, T^{\nu\, (0)}_{\mu}
\label{eq:einstein5},
\end{equation}
where the gravitational coupling constant $G$ can be regarded as the expansion parameter and the linearized Einstein tensor is given by
\begin{equation}
G^{\nu\, (1)}_{\mu} = g^{\nu \rho}R^{(1)}_{\mu \rho}
-h^{\nu \rho}R^{(0)}_{\mu\rho}-\frac{1}{2}\delta_{\mu}^{\nu}
\Big(R^{\alpha\, (1)}_\alpha -h^{\alpha\beta}R^{(0)}_{\alpha\beta}\Big)
\label{eq:einsteinT},
\end{equation}
with $R^{(1)}_{\mu \nu}$ given by eq.~\eqref{eq:einstein5} and $R^{(0)}_{\mu\nu} = 3 g_{\mu\nu}$ being the Ricci tensor of the de Sitter background. Note that when perturbing the Einstein equation with one of the indices raised, the contribution of the cosmological constant term is entirely contained in the background equation and it does not appear in eq.~\eqref{eq:einstein5} for the linear perturbations.
As for the background stress tensor $T^{\nu\, (0)}_{\mu} = g_{\mu\rho} T_{(0)}^{\rho\nu}$ of a free-falling point-like mass, one can take
\begin{equation}
T_{(0)}^{\mu\nu} = m \int d\tau \, \frac{\dot{z^\mu} (\tau) \dot{z^\nu} (\tau)}
{\sqrt{-\dot{z}^2} \sqrt{-g}} \, \delta^{(4)} \big(x^\alpha - z^\alpha(\tau)\big)
= m \frac{\dot{z^\mu} \dot{z^\nu}}{\sqrt{g^{(3)}}} \,
\delta^{(3)} \big(x^\alpha - z^\alpha(\tau)\big)
\label{eq:stress1},
\end{equation}
where we have used the same notation as in eq.~\eqref{eq:current1} and $\dot{z}^2 \equiv g_{\mu\nu} \dot{z}^\mu \dot{z}^\nu$.

In order to find an explicit solution in spatially flat coordinates, we will make use of the results derived in Ref.~\cite{deVega}. Defining $\psi^{\mu}_{\nu} \equiv h^{\mu}_{\nu} - (h/2) \delta^\mu_{\nu}$ and imposing the gauge condition $\nabla_\mu \psi^{\mu}_{\nu} = - (2/\eta) \psi_{\nu}^{0}$, a closed set of equations governing the dynamics of $\chi_{\mu\nu}=\eta_{\nu\alpha}\psi^\alpha_\mu$ were obtained in that reference. One can study the solutions of the inhomogeneous and homogeneous equation. The homogenous equation satisfied by the two physical polarizations of the gravitational waves correspond to the Klein-Gordon equation for a massless minimally coupled scalar field. Here we will, however, focus on the solutions of the inhomogeneous equation. For a point-like mass located at $x^i = 0$, the only non-vanishing component that results from evaluating eq.~\eqref{eq:stress1} is $T^{(0)}_{00} = m (-\eta)\, \delta^{(3)} (x^i)$, where we took into account that $\dot{z}^\mu = e^{-\tau} \delta^\mu_0 = -\eta\, \delta^\mu_0$ in the $\{\eta,x^i\}$ coordinates that we are employing.
The nontrivial part of the Einstein equations is then
\bqa
\partial^2\chi_{00}+\frac{2}{\eta}\partial_{0}\chi_{00}-\frac{2}{\eta^2}\chi_{00}
&=& -16\pi G\, T^{(0)}_{00} \nonumber \\
\mathrm{or,\  equivalently,} \quad \partial^2 \! \left(\frac{\chi_{00}}{\eta}\right)
&=& -16\pi G\, \frac{T^{(0)}_{00}}{\eta}
\label{eq:nontriv},
\eqa
where $\partial^2 = \eta^{\mu\nu} \partial_\mu\partial_\nu$ is just the flat space D'Alembertian. A particular solution of eq.~\eqref{eq:nontriv} is the following:
\begin{equation}
\frac{\chi_{00}}{\eta} =  \frac{4G m}{|\vec{x}|}
\quad \Rightarrow \quad
\psi^0_0=-\chi_{00} = - \frac{4G m\,\eta}{|\vec{x}|},
\end{equation}
which implies
\begin{equation}
h^0_0 = - \frac{2 G m\, \eta}{|\vec{x}|},
\qquad h^j_i = \frac{2 G m\, \eta}{|\vec{x}|}\, \delta_i^j
\label{eq:perturbation1}.
\end{equation}
Following Ref.~\cite{weinberg72} we will consider $t^{\mu}_{\nu} = (1/8\pi G)\, G^{\mu\, (2)}_{\nu}$ as the stress tensor of the linearized gravitational field. It can be directly obtained by evaluating the exact Einstein tensor for the perturbed metric $\tilde{g}_{\mu\nu}$ [using the result in eq.~\eqref{eq:perturbation1}] and keeping only the terms quadratic in the gravitational coupling constant $G$. In this way we find
\begin{equation}
t^{0}_{i}\propto\frac{Gm\,x_{i}\,\eta^3}{|\vec{x}|^4} ,
\end{equation}
which gives a total integrated flux over the sphere that falls off to zero at large radii.

\subsection{Static coordinates}

The perturbative solution for a point-like particle at $r=0$ in static coordinates can be obtained by expanding perturbatively in $G$ the exact Schwarzschild-de Sitter solution
\bqa
ds^2 = -\left(1-\frac{2\,G m}{r} - r^2\right)dt^2 + \frac{dr^2}{\left(1-\frac{2\,G m}{r} - r^2\right)} + r^2\, d\Omega_2^2  \label{eq:schwarzschild_dS}.
\eqa
Keeping only terms linear in $G$ one gets the following result for the linear metric perturbations:
\begin{equation}
h_{tt} = \frac{2 G m}{r},
\qquad h_{rr}= \frac{2 G m}{r(1-r^2)^2}
\label{eq:perturbation2}.
\end{equation}
Now we should use this metric perturbation to construct $\tilde{g}_{\mu\nu}$ and calculate $t^{\mu}_{\nu} = (1/8\pi G)\, G^{\mu\, (2)}_{\nu}$. Alternatively one can consider the exact Einstein tensor for the full Schwarzschild-de Sitter solution and keep those terms quadratic in $G$. In fact, since the Schwarzschild-de Sitter solution satisfies the equation $G_\mu^\nu = 3\, \delta_\mu^\nu$ (remember that $\Lambda=3$ in our units), there is no dependence on $G$ left when evaluating the full Einstein tensor outside the sources. This will, however, include a possible contribution to the full Einstein tensor from terms quadratic in $G$ in the exact Schwarzschild-de Sitter metric, which needs to be subtracted. Therefore, one needs to take $h_{\mu\nu}^{(2)} = 4 G^2m^2/r^2(1-r^2)^3\, \delta_\mu^r \delta_\nu^r$ and use eq.~\eqref{eq:lin_ricci1} to calculate its contribution to the perturbed Ricci tensor, and from it get immediately the Einstein tensor contribution to be subtracted. Fortunately, one can easily see that the $0i$ components will vanish and we can conclude that $t^{0}_{i} = 0$, as one intuitively expects from the existence of a static solution.

\subsection{Global coordinates}
\label{sec:gravity_global}

Similarly to the electromagnetic case, one also needs to consider at least two sources in the case of linearized gravity. Analogously to the requirement of a vanishing total electric charge implied by the Gauss constraint, the constraints from the Einstein equations impose the following restriction on the lowest-order stress tensor of the sources:
\begin{equation}
\int_{\Sigma_t} d^3 x \sqrt{g_{_\Sigma}} \, T_{\mu\nu}^{(0)}\, n^\mu \, \xi^\nu = 0 ,
\end{equation}
given a compact Cauchy hypersurface $\Sigma_t$ with normal unit vector $n^\mu$, and a Killing vector $\xi^\nu$. (It is crucial to have a compact Cauchy hypersurface; that is why this issue arises in full de Sitter space, whose Cauchy hypersurfaces have an $S^3$ topology, but not in the subregion covered by spatially flat coordinates.)
This restriction is closely related to the phenomenon of \emph{linearization instability} \cite{brill73} in this context and is sometimes referred to as ``linearization instability constraint''. In particular, for the timelike Killing vectors in de Sitter this condition is not satisfied by a single particle, but it is fulfilled by two antipodal free-falling particles with equal mass. In fact, the simplest way to obtain the linearized gravity solution for this case is to consider the small mass limit of the Schwarzschild-de Sitter solution, restrict oneself to the region outside the black-hole horizon (including the extension beyond the cosmological horizon) and treat the mass perturbatively there. In this way, we can directly obtain the metric perturbation in terms of the static background coordinates from eq.~\eqref{eq:schwarzschild_dS}.
The result (obtained in the previous subsection) will cover the two antipodal static patches of de Sitter, but it can be regularly extended to the full de Sitter spacetime by considering the extension of the static coordinates to $r>1$ as described in Appendix~A.3, and even changing to global background coordinates using eqs.~\eqref{eq:transf3}-\eqref{eq:transf5}.

%@@ It would be nice to check that this solution coincides with that obtained
%using spatially flat coordinates. A simple way to do that would be to compute
% the linearized Weyl tensor in both cases, express them in the same background
%coordinates and compare them. @@

\section{Late-time results for an arbitrary set of geodesics}
\label{sec:late}

At sufficiently late times a general set of geodesics in de Sitter spacetime will become arbitrarily far apart. We will argue that in most cases the field generated by a general set of free-falling charges moving along such geodesics will exhibit a generic form at sufficiently late times.

\subsection{Scalar field}

The basic idea is simple. We saw in Sec.~\ref{sec:general_scalar2} that for $M^2>0$ the field generated by a single free-falling charge tends at late times to a generic solution $\phi_\mathrm{s}(\tau,x^i)$, which in turn falls off at large distances. The generalization to multiple charges is straightforward: since we are considering linear fields, one can simply apply the superposition principle.
(One would need external forces to compensate the mutual forces and guarantee that the charges continue to follow geodesics, at least until the mutual forces can be neglected, which is the late-time regime that we are really interested in here.)
Furthermore, at sufficiently late times so that all the geodesics are sufficiently far apart [and taking into account the fall-off of $\phi_\mathrm{s}(\tau,x^i)$ at large distances], one can focus entirely on the field generated by the closest charge and neglect the effect from all the other charges.

Therefore, the generic late-time field configuration would correspond to effectively non-overlapping field zones surrounding each one of the charges, where one could employ adapted spatially flat and even static coordinates (for points within the horizon of the charge) such that the results of Secs.~\ref{sec:general_scalar1}-\ref{sec:general_scalar2} would directly apply. This result, however, does not hold for $M^2=0$, as we found in Sec.~\ref{sec:minimal} when studying explicitly the massless minimally coupled case. On the other hand, for $0 < M^2 \ll 1$ the conclusion will be true, but one will need to wait for very long times to make sure that the solution associated to each one of the charges decays to $\phi_\mathrm{s}(\tau,x^i)$ and that the different charges are separate enough. This is because $\lambda_-$, which controls the decay rate of any homogenous solution of the Klein-Gordon equation as well as the fall-off of $\phi_\mathrm{s}(\tau,x^i)$ with the distance, is very small in that case.

%@@ Note also that although we argued in terms of spatially flat coordinates,
%the same conclusions would hold if we had considered global coordinates.
%full dS... antipodal @@

\subsection{Electromagnetic field}

The situation for the {electromagnetic field} is similar to that for scalar fields with $M^2 > 0$. We saw in Sec.~\ref{sec:e-m} that there is a static solution in the static patch which can be regularly extended to the full spacetime. The remaining question is then whether the solutions of the homogenous Maxwell equation decay at late times analogously to the solutions of the Klein-Gordon equation. This can be easily investigated by taking advantage of the invariance of Maxwell's equation under conformal transformations. For instance, in spatially flat coordinates the solutions for $F_{\mu\nu}$ are the same as in flat space, i.e.\ free electromagnetic waves. However, the physical electromagnetic field (referred to an orthonormal basis) should be rescaled by the inverse square of the scale factor: $F_{\hat{\mu}\hat{\nu}} = \eta^2 F_{\mu\nu}$. Physically this corresponds to the amplitude of the electromagnetic waves being redshifted away as they propagate in the expanding spacetime, and results in any initial solution decaying to that of Sec.~\ref{sec:e-m} at sufficiently late times, or the corresponding generalization for an arbitrary set of free-falling charges analogous to that described for scalar fields in the previous subsection.

\subsection{Linearized gravity}

Things are slightly more subtle for {linearized gravity}. As seen in Sec.~\ref{sec:gravity_global}, there is a static solution which can be regularly extended to the full spacetime. However, whether other solutions generically decay to it at late times is less obvious. After an appropriate gauge-fixing and field redefinition, one can see that the homogeneous equation satisfied by the two physical polarizations of the linear metric perturbations are that of a massless minimally coupled scalar field \cite{Ford:1977dj}. This would seem to imply that given a non-vanishing initial condition, it will not decay at late times.
Nevertheless, one should consider gauge invariant quantities which naturally describe the local geometrical properties.
%Nevertheless, one should bear in mind that only gauge invariant quantities can
%provide meaningful information free of spurious effects.
The linearized Weyl tensor, $C^{(1)}_{\mu\nu\rho\sigma}$, is one such object which fully characterizes the perturbed geometry to linear oder. One can check that the linearized Weyl tensor for the homogeneous solutions of the linearized Einstein equation does decay at late times, so that the static solution and its extension indeed correspond to an attractor geometry at late times.

\section{Discussion}
\label{sec:discussion}

We have analyzed the classical fields generated by free-falling charges in de Sitter spacetime coupled to different kinds of fields. We were able to provide explicit results for the electromagnetic case, linearized gravity, and scalar fields with a wide range of values for the mass and the curvature coupling ($M^2=m^2 +\xi R \geq 0$). Furthermore, we have argued that at sufficiently late times one should have a generic result for an arbitrary set of free-falling charges in all cases (other than a scalar field with $M^2=0$), which would correspond to each charge being surrounded by a field zone with negligible influence from the other charges and where the results obtained for a single charge would apply.

Except for scalar fields with $M^2=0$ we have generically found ``absence of radiation'' in the static patch associated with each free-falling charge. More precisely, we have obtained a static solution with vanishing energy flux with respect to this frame. This implies conservation of the Killing energy and, together with the static character of the field, can be exploited to argue that there was no energy loss (emission) by the source. As discussed in Sec.~\ref{sec:general_scalar1} and Appendix~B, having a static solution in the static patch is simple, the non-trivial aspect is whether it is regular on the horizon so that it can be extended to the full de Sitter spacetime.

Furthermore, we have also studied the energy flux at distances much larger than the horizon as seen by the natural set of inertial observers in spatially flat coordinates, and found a remarkable result. For sufficiently light fields (with $M^2 < 5H^2/4$) the energy flux does not fall off fast enough (slower than the inverse square of the distance) and the total flux integrated over a sphere of radius $R$ becomes arbitrarily large for large radii. This phenomenon can, in fact, be qualitatively understood as the result of a classical parametric amplification process. In contrast with the Killing energy in the static patch, there is no energy conservation in the natural frame associated with the spatially flat coordinates (and in any case there is no frame where energy is conserved for scales larger than the horizon), and one can interpret that the field generated by the free-falling charge is being amplified by the time-dependent background spacetime, analogously to what would happen to a harmonic oscillator with a time dependent frequency (e.g. a pendulum with a time-dependent length).

Note also that the unbounded growth of the integrated flux with the radius does not necessarily entail dramatic effects for two reasons. First, the stress tensor itself is small (the flux component is quadratically suppressed by the mass of the light field), so that its contribution to the back-reaction on the spacetime geometry will be locally small (although a more detailed analysis would be required to make sure that there are no significant secular effects). Second, while this kind of phenomena would face an insurmountable tension with energy conservation in flat spacetime, it becomes much easier to avoid contradictions in de Sitter, where energy is not globally conserved.

On the other hand, for scalar fields with $M^2=0$ there is no stationary solution in the static patch which can be regularly extended to the full spacetime. Instead we explicitly found for the massless minimally coupled case that the charge loses energy at a constant rate. As discussed in Sec.~\ref{sec:minimal_static}, this means that its effective mass will vanish and start to become negative within a finite time. In Ref.~\cite{Poisson} it was assumed that the charge would simply disappear at that point. However, as illustrated in Sec.~\ref{sec:minimal_static} with the example of a particle which is also electrically charged, we believe that the question can be more subtle and should be studied within an appropriate microscopic model of the charges. It seems clear that a negative mass signals some kind of instability, but its precise nature is less obvious. For instance, if one tried to model the microscopic constituents of the charge by massless fermions with a Yukawa coupling to the field, it is known that the theory exhibits an instability unless there is a minimum amount of nonlinear self-coupling for the scalar field. It might be that for certain parameter ranges the theory itself is unstable in de Sitter space, or that the effects of the non-linear terms in the scalar field potential need to be taken into account when considering the late-time evolution of the effective mass.

\begin{acknowledgments}
E.T.A.\ would like to acknowledge discussions with O.~Kancheli and Y.~Khriplovich. E.T.A.\ would also like to thank the Albert-Einstein-Institut, where most of this work was carried out, for hospitality.
The work of E.T.A.\ was done under the partial financial support of grants for the Leading Scientific Schools NSh-6260.2010.2 and RFBR 08-02-00661-a.
\end{acknowledgments}

\appendix

\section*{APPENDIX A: COORDINATE SYSTEMS IN DE SITTER SPACE}
\label{appA}

Four-dimensional de Sitter space with unit radius of curvature ($H=1$) can be defined as the hyperboloid given by $\eta_{AB} X^A X^B = 1$ and embedded in the five-dimensional Minkowski spacetime with metric $\eta_{AB} = \mathrm{diag}(-1,1,1,1,1)$. It is invariant under the ten-dimensional $O(4,1)$ group of isometries, which corresponds to the Lorentz group of the embedding spacetime.

\subsection{Global coordinates}

One can introduce a coordinate system $\{T,\theta_i\}$ which covers the whole hyperboloid and corresponds to the following embedding:
\bqa
X^0 &=& \sinh(T),\quad
X^a = - \cosh(T)\,\omega^a, \quad a=1,2,3,4,\nonumber \\
\omega^1 &=& \sin(\theta_3)\,\sin(\theta_2)\,\sin(\theta_1),\nonumber \\
\omega^2 &=& \sin(\theta_3)\,\sin(\theta_2)\,\cos(\theta_1),\nonumber \\
\omega^3 &=& \sin(\theta_3)\,\cos(\theta_2), \nonumber \\
\omega^4 &=& \cos(\theta_3)
\label{eq:global0},
\eqa
where $\delta_{ab} \omega^a \omega^b =1$ and defines a 3-sphere.
The line element in these coordinates is
\bqa
ds^2 = -dT^2 + \cosh^2(T)\, d\Omega_3^2
= \frac{1}{\cos^2(\lambda)}\, \left[-d\lambda^2 + d\Omega_3^2\right]
\label{eq:global1},
\eqa
with $d\Omega_3^2 = d\theta_3^2 + \sin^2(\theta_3)\,\left[d\theta_2^2 + \sin^2(\theta_2)\, d\theta_1^2\right]$. The last equality in eq.~\eq{eq:global1} shows explicitly that de Sitter space is conformal to a region of the Einstein static universe characterized by the finite range of the conformal time, $-\pi/2 < \lambda < \pi/2$. Six of the ten independent Killing vectors are manifest in these global coordinates, namely, those associated with the $O(4)$ isometry group of the 3-spheres that correspond to the spatial sections with $T=\mathrm{const}$.

The worldlines with $\theta_i=\mathrm{const.}$ are timelike geodesics with proper time $T$ and can be regarded as inertial, or free-falling, observers. (Note incidentally that any pair of geodesics can be related by a de Sitter isometry transformation.) Throughout the text we often consider the particular geodesic defined by $\theta_3=0$, and the antipodal one, with $\theta_3=\pi$.

\subsection{Spatially flat coordinates}

The following alternative parameterization covers half of the hyperboloid, corresponding to $X^0 - X^4 > 0$ (the other half can be covered by an analogous parameterization which changes the sign of the $X^0$ and $X^4$ coordinates):
%% This half corresponds to $X^0 - X^4 > 0$.
\bqa
X^0 &=& - \frac{1+\vec{x}^2}{2\,\eta} + \frac{\eta}{2}\nonumber \\
X^i &=& - \frac{x^i}{\eta}, \quad i=1,2,3\nonumber \\
X^4 &=& \frac{1-\vec{x}^2}{2\,\eta} + \frac{\eta}{2} \label{eq:planar1},
\eqa
with $\eta \in (-\infty,0)$. The line element can then be written as
\begin{equation}
ds^2= -d\tau^2 + e^{2\tau} \delta_{ij} dx^i dx^j
= \frac{1}{\eta^2}(-d\eta^2 + \delta_{ij} dx^i dx^j) \label{eq:planar2},
\end{equation}
where the comoving time $\tau$ is related to the conformal time $\eta$ by $\eta = -\exp(-\tau)$. In terms of the conformal coordinates $\{\eta,x^i\}$ it is obvious that this half of de Sitter is conformal to half of Minkowski space with $\eta \in (-\infty,0)$. Furthermore, seven of the isometry generators of de Sitter space are manifest in this coordinate system. In addition to the six generators of the Euclidean group $E(3)$ for the spatial sections with $\eta=\mathrm{const.}$, the metric in eq.~\eq{eq:planar1} is manifestly invariant under the following transformation:
\bqa
x^i &\to& e^{-\alpha} x^i , \nonumber\\
\eta  &\to& e^{-\alpha} \eta
\qquad \mathrm{or,\ equivalently,} \qquad \tau \to \tau + \alpha
\label{eq:isometry1}.
\eqa
%or, equivalently, $\tau \to \tau + \alpha$.

The worldlines with $x^i = \mathrm{const.}$ are timelike geodesics with proper time $\tau$, and here the particular geodesic corresponding to $\theta_3=0$ in the previous subsection is given in conformal coordinates by $x^\mu(\tau)=(- e^{-\tau},0,0,0)$. In contrast, the antipodal geodesic located at $\theta_3=\pi$ is entirely contained in the other half of de Sitter space not covered by this parameterization.

\subsection{Static coordinates}

While de Sitter spacetime has no global timelike Killing vectors, it does have Killing vectors which are timelike in half of the spacetime and exhibit a bifurcate Killing horizon structure. One can introduce static coordinates which cover each one of the two patches where the Killing vector is timelike (hence one quarter of the full de Sitter space) and correspond to the following parameterization of the hyperboloid:
\bqa
X^0 &=& \sqrt{1-r^2} \, \sinh(t),\nonumber \\
X^i &=& - r \, \omega^i, \quad i=1,2,3,\nonumber \\
X^4 &=& - \sqrt{1-r^2} \, \cosh(t) \label{eq:static1},
\eqa
where $0 \leq r < 1$ and $\omega^i$, which satisfy the condition $\delta_{ij} \omega^i \omega^j =1$ and define a 2-sphere, can be parametrized as
\bqa
\omega^1 &=& \sin(\theta_2)\,\sin(\theta_1),\nonumber \\
\omega^2 &=& \sin(\theta_2)\,\cos(\theta_1), \nonumber \\
\omega^3 &=& \cos(\theta_2).
\eqa
The line element in these coordinates is given by
\bqa\label{staticor}
ds^2 = - (1 - r^2)\, dt^2 + \frac{dr^2}{(1 - r^2)} + r^2 d\Omega_2^2
\label{eq:static2},
\eqa
where $d\Omega_2^2 = d\theta_2^2 + \sin^2(\theta_2)\, d\theta_1^2$.
Four independent isometry generators are manifest in these coordinates: three associated with the $O(3)$ group of the 2-sphere and one corresponding to the time-translation symmetry $t \to t + \alpha$.
The worldlines of constant $r$ are uniformly accelerated timelike curves and only $r=0$ is a geodesic (with proper time $t$), which coincides with the same geodesic considered in the previous two subsections; $r=1$ corresponds to the \emph{event horizon} surrounding this geodesic observer.

Note that there is an analogous static patch, antipodal to the one considered so far, which can be covered by the same kind of coordinates by changing the sign of both $X^0$ and $X^4$ in the above parameterization. Furthermore, the remaining two quadrants of de Sitter can be covered by extending the range of the radial coordinate to $1<r<\infty$ and making the following changes in the embedding parameterization described by eq.~\eq{eq:static1}:
\bqa
X^0 &=& \pm \sqrt{r^2-1} \,\cosh(t), \nonumber \\
X^4 &=& \mp \sqrt{r^2-1} \, \sinh(t) \label{eq:static3}.
\eqa
The two additional quadrants covered by this parameterizations are not static since the timelike and spacelike character of $\partial_t$ and $\partial_r$ are interchanged.

We conclude this appendix briefly describing  the relation between the static coordinates (as well as their extensions) and the coordinate systems of the previous subsections.

\subsubsection{Relation to spatially flat coordinates}

Comparing the expressions for $X^i$ in eqs.~\eq{eq:planar1} and \eq{eq:static1} one immediately finds
\begin{equation}
r = -\frac{|\vec{x}|}{\eta} = a(\eta)\, |\vec{x}| \label{eq:transf1}.
\end{equation}
Similarly, by comparing the expressions for $X^0 - X^4$, one gets
\begin{equation}
-\frac{1}{\eta} =  e^t \, \sqrt{1-r^2} \quad \mathrm{or} \quad
e^{-2t} = \eta^2 - |\vec{x}|^2 \label{eq:transf2}.
\end{equation}
Taking the logarithm, the following relation between the comoving time of spatially flat coordinates and the static time is obtained:%
\footnote{If one extends the $\{r,t\}$ coordinates outside the static patch using eqs.~\eq{eq:static3}, there the relation becomes $\tau = - t + (1/2) \ln (r^2-1)$.}
\begin{equation}
\tau = t + \frac{1}{2} \ln (1-r^2) .
\end{equation}
They both coincide along the worldline at $r=0$ and correspond to the proper time of that geodesic. Furthermore, from these relations it is straightforward to see that the isometry transformation defined in eqs.~\eq{eq:isometry1} corresponds to the time translation symmetry for $t \to t + \alpha$ keeping $r$ fixed in static coordinates.

\subsubsection{Relation to global coordinates}

Finally, comparing $X^i$ (with $i=1,2,3$) as well as $X^0/X^4$ in eqs.~\eq{eq:global0} and \eq{eq:static1}, one finds the following relation between global and static coordinates:
\bqa
r &=& \cosh(T) \sin(\theta_3), \label{eq:transf3} \\
\tanh (t) &=& \frac{\tanh (T)}{\cos(\theta_3)} \label{eq:transf4}.
\eqa
If one extends the static coordinates outside the static patch, the last relation becomes
\begin{equation}
\coth (t) = \frac{\tanh (T)}{\cos(\theta_3)} \label{eq:transf5}.
\end{equation}

\setcounter{subsection}{0} %% !!!

\section*{APPENDIX B: REGULAR EXTENSION OF THE STATIC SOLUTION}
\label{appB}

The static solution for a free-falling charge in de Sitter found in Sec.~\ref{sec:general_scalar} is directly related to a solution of the Euclidean version of the Klein-Gordon equation on the 4-sphere. This fact can be exploited to show that for $M^2>0$ the solution regular on the horizon can always be matched to the solution with a source at $r=0$. This will be shown in the first subsection of this appendix by arguing that if the contrary were true, there would be a solution regular everywhere on the sphere, and then proving that this is impossible for $M^2>0$. In the second subsection we will discuss the extension of this solution to the full de Sitter spacetime and its regularity.

\subsection{Equivalent problem on the 4-sphere}

A 4-sphere can be obtained by analytic continuation from de Sitter space by taking $t =\imath \theta$ in eq.~\eq{eq:static2} and requiring $\theta$ to have a period of $2\pi$ so that there is no conical singularity at $r=1$. This becomes perhaps even clearer if we introduce the change of coordinates $r=\sin(\chi)$, with $0\leq\chi\leq\pi$, so that the line element becomes
\begin{equation}
ds^2 = \cos^2(\chi) d\theta^2 + d\chi^2 + \sin^2(\chi) d\Omega_2^2
\label{eq:sphere1}.
\end{equation}
One can proceed in the same way to obtain the Euclidean version of the Klein-Gordon equation~\eq{eq:static5}. In the new variables the source term is given by $(q/4\pi \chi^2) \delta(\chi) + \big(q/4\pi (\chi-\pi)^2\big) \delta(\chi-\pi)$, where the second term would correspond to the second static patch mentioned in Appendix~A.3 and is required here for continuity in $\theta$. If one looks for solutions of the form $\phi(\chi)=\phi(r)$ (in fact, for $M^2\geq0$ they are necessarily of this form), the equation satisfied becomes equivalent to eq.~\eq{eq:static6}.%
\footnote{Note that it makes sense to identify $\phi(\chi)=\phi(r)$ only if $\phi(\chi)=\phi(\pi-\chi)$. This is possible because the smooth function that we will find below is an even function of $(\chi-\pi/2)$. Hence, from now on we will focus on $\chi\leq\pi/2$ and use $\phi(\chi)=\phi(\pi-\chi)$ for $\chi\geq\pi/2$.}
Thus, one can make use of the results in eqs.~\eq{eq:frobenius1a}-\eq{eq:frobenius2b} for the expansions around the regular singular points at $\chi=0$ and $\chi=\pi/2$ (corresponding respectively to $r=0$ and $r=1$).
Taking into account that $r=\sin(\chi)$, it is clear that $\phi_1(\chi)$ from eq.~\eq{eq:frobenius1a} is an even function of $\chi$ with no spike at $\chi=0$, where it is smooth. Similarly, taking into account that $1-r = 1 - \cos\big(\chi-\pi/2\big)$, one can see that $\tilde{\phi}_1(\chi)$ from eq.~\eq{eq:frobenius2a} is an even function of $(\chi-\pi/2)$ with no spike at $\chi=\pi/2$, where it is smooth.

The advantage of working on the 4-sphere is that we can easily prove that for $M^2>0$, $C \tilde{\phi}_1(\chi)$ always matches a linear combination $A \phi_1(\chi) + B \phi_2(\chi)$ with non-vanishing $B$. Indeed, if $B$ vanished, $C \tilde{\phi}_1(\chi)$ would match $A \phi_1(\chi)$ and would correspond to a homogeneous solution of eq.~\eq{eq:static6} which would be regular everywhere on the 4-sphere, but one can show that this is impossible as follows. The homogeneous solutions of eq.~\eq{eq:static6} are solutions of the Euclidean version of the Klein-Gordon equation:
\begin{equation}
(\Delta_4 - M^2)\, \phi = 0
\label{eq:sphere2},
\end{equation}
where $\Delta_4$ is the Laplace operator on $S^4$, whose eigenvalues are $-L(L+3)$ with $L=0,1,\ldots$ Therefore, one cannot have a regular solution for $M^2>0$ since that would correspond to an eigenfunction of the Laplacian with a positive eigenvalue. In contrast, for $M^2=0$ one does have the obvious solution $\phi = \mathrm{const.}$, which explains why our argument does not work for the massless minimally coupled case and why one cannot find a static solution regular on the horizon associated with a freely-falling charge in that case.

\subsection{Extension to full de Sitter spacetime}

One can obtain de Sitter space by analytic continuation from $S^4$ reversing the procedure at the beginning of the previous subsection. In that way, one is naturally led to a static patch of de Sitter, but one can go to full de Sitter by extending the static coordinates to $r>1$, as explained in Appendix~A.3. Alternatively, one can consider an analytic continuation of the form $\imath T = \theta_4 - \pi/2$ to the global coordinates. On the 4-sphere the source is located on an equatorial worldline with $\theta_3=\{0,\pi\}$ and arbitrary values of $\theta_4$, which becomes two disconnected antipodal worldlines in de Sitter located respectively at $\theta_3=0$ and $\theta_3=\pi$, and arbitrary values of $T$.

Since we found a function which is regular everywhere on $S^4$ (outside the source), the analytic continuation to full de Sitter space should be regular too. This can be explicitly checked by using eq.~\eq{eq:transf3} to express in terms of global coordinates the solution $\phi_\mathrm{s}(r)$ found in Sec.~\ref{sec:general_scalar1}. The only potentially problematic value is $r=1$ and, as seen in eq.~\eq{eq:frobenius2a}, the solution around that value can be expanded as a convergent series in powers of $(1-r)$, which can be written in global coordinates as
\begin{equation}
1-r = 1 - \cosh(T) \sin(\theta_3) = 1 - \frac{\sin(\theta_3)}{\cos(\lambda)}
=  1 - \frac{\cos(\theta_3-\pi/2)}{\cos(\lambda)}
\label{eq:extension1},
\end{equation}
where the neighborhood of $r=1$ can be more easily studied in terms of the conformal time $\lambda$, since $r=1$ then corresponds to $\theta_3 - \pi/2 = \pm \lambda \mod \pi$. The solution is, thus, manifestly smooth everywhere outside the sources. Note that in order not to have a spike at $\theta_3-\pi/2=\lambda=0$ it is important that $\phi_\mathrm{s}(r)$ is an even function of $\theta_3-\pi/2$ and $\lambda$, which follows from eq.~\eq{eq:transf3} and the fact that the solution in eq.~\eq{eq:frobenius2a} is analytic in $(1-r)$.

Besides the neighborhood of $r=1$ one can also study the behavior at late times (and fixed comoving distance) or large physical distance from the sources (and fixed time) by considering the solution $\phi_\mathrm{s}(r)$ for large values of $r$. More specifically, one can introduce the change of variables $w=1/r$ in eq.~\eq{eq:static6} (note that for large $r$ the source is irrelevant) and see that $w=0$ is a regular singular point. One can, therefore, make use of the Frobenius method and Fuchs's theorem to express the solutions as a series expansion in powers of $1/r$, analogously to what was done in Sec.~\ref{sec:general_scalar1}. The two possible solutions are given by
\begin{eqnarray}
\hat{\phi}_1(r) &=&  r^{-\lambda_-} \, \big(1+\cdots\big) \,, \label{eq:frobenius3a}\\
\hat{\phi}_2(r) &=& r^{-\lambda_+} \, \big(1+\cdots\big)  \,, \label{eq:frobenius3b}
\end{eqnarray}
where $\lambda_\pm = 3/2 \pm \sqrt{9/4-M^2}$ (notice that for $M^2>0$ their real part is always positive) and the remaining terms involve positive integer powers of $1/r$ times a possible $\ln r$ factor in some cases.
Using $r = \cosh(T) \sin(\theta_3)$ one can straightforwardly obtain the late-time and large-distance behaviors in global coordinates. (The late-time behavior coincides with that obtained in Sec.~\ref{sec:general_scalar2} for the homogeneous solutions of the Klein-Gordon equation because the source played no role here.) The extension of the static solution will correspond to a unique linear combination $D\, \hat{\phi}_1(r) + E\, \hat{\phi}_2(r)$ whose coefficients should be determined by matching it to the regular solution near the horizon, given by eq.~\eq{eq:frobenius2a}, with the right normalization factor. This would require solving exactly \eq{eq:static6}. Lacking an exact solution, we can still expect that the large $r$ behavior will be determined by $\hat{\phi}_1(r)$, which dominates over $\hat{\phi}_2(r)$ at large $r$, as long as $D\neq0$ (that this is the case could perhaps be shown using approximate methods to match the solutions). For the particular example of a conformal field (corresponding to $M^2=2$), we know that the exact solution is proportional to $1/r$ and $D\neq0$ indeed (while $E=0$).

\end{document}